\documentclass[10pt,twocolumn]{IEEEtran}
\usepackage{amsmath,amscd,amssymb,amsgen,amsfonts,amsbsy}
\usepackage{mathrsfs}
\usepackage[utf8x]{inputenc}
\usepackage{color}
\usepackage{textcomp}
\usepackage{float}
\usepackage{latexsym,graphicx}
\usepackage{tikz}
\usetikzlibrary{automata,arrows,positioning,calc}
\usepackage{url}
\usepackage{algorithmic}
\usepackage[ruled,lined]{algorithm2e}

\newcommand{\prob}[1]{\mathsf{Pr}\left( #1 \right)}
\newcommand{\expect}[1]{\mathsf{E}\left({#1}\right)}

\newcommand{\remove}[1]{}

\newcommand{\comments}[1]{}

\newcommand{\qed}{\hfill $\square$}

\newtheorem{corollary}{Corollary}
\newtheorem{lemma}{Lemma}
\newtheorem{proposition}{Proposition}
\newtheorem{property}{Property}
\newtheorem{theorem}{Theorem}
\newtheorem{remark}{Remark}
\newtheorem{definition}{Definition}

\newtheorem{assumption}{Assumption}
\makeatother

\title{Simulation Based  Algorithms for Markov Decision Processes and Multi-Action Restless  Bandits}

\author{Rahul Meshram and Kesav Kaza \thanks{Rahul Meshram is with the Electrical Engineering Department, IIT Madras, Chennai, India.  Email: rahulmeshram07@gmail.com. Kesav Kaza is with the Electrical Engineering Department, IIT Bombay, Mumbai, India.  Email: kesav.kaza@gmail.com. }}
\begin{document}
\maketitle

\begin{abstract}
	
	We consider multi-dimensional Markov decision processes and formulate a long term  discounted reward optimization problem.  Two simulation based algorithms---Monte Carlo rollout policy and parallel rollout policy are studied, and various properties for these policies are discussed.  
	
	We next consider a restless multi-armed bandit (RMAB) with multi-dimensional state space and multi-actions bandit model.  A standard RMAB consists of two actions for each arms whereas in multi-actions RMAB, there are more that two actions for each arms.
	A popular approach for RMAB is  Whittle index based heuristic policy. Indexability is an important requirement to use index based policy. Based on this, an RMAB is classified into indexable or non-indexable bandits.  Our interest is in the study of Monte-Carlo rollout policy for both indexable and non-indexable restless bandits. We first analyze a standard indexable RMAB (two-action model) and discuss an index based policy approach. We present approximate index computation algorithm using Monte-Carlo rollout policy. This algorithm's convergence is shown using  two-timescale stochastic approximation scheme.  Later, we analyze multi-actions indexable RMAB, and discuss the index based policy approach. We also study non-indexable RMAB for both standard and multi-actions bandits using Monte-Carlo rollout policy.  
	
\end{abstract}

\section{Introduction}
%\rahul{How are you going to explain the ideas in the paper, what are ideas you are going to explain, why are they important}
Markov decision processes have been extensively studied in the literature for various applications, e.g. wireless communication systems and networks \cite{Alsheikh15}, internet of things (age of information) \cite{Hsu17} and queueing systems for cloud management \cite{NinoMora19}. Markov decision processes are a class of sequential decision problems--- an environment is modeled using a state variable, the decision maker observes the current state, acts on the environment by taking a certain action, and the environment reacts by changing state. The environment changes state according to some probability law, \cite{Puterman14,Bertsekas95,Bertsekas96}. The decision maker (DM) obtains a reward (cost) from a given state based action (decision). The objective of the decision maker is choose actions in sequence such that it maximizes (minimize) long term expected reward (cost). 

Most often, solution methodologies used for MDP are 1) value iteration algorithm, and 2) policy iteration algorithm, \cite{Puterman14,Bertsekas95}.  In both algorithms, DM needs to know the state transition probabilities and rewards. Further, in order to implement these algorithms in practice, MDP is assumed to have finite state and  action spaces. Then, numerical computation can be performed using these algorithms and  optimal long term reward can be obtained. 

In this paper, we consider multi-dimensional Markov decision processes, where complexity of MDP can be due to 1) large number of states (countable or uncountable), 2) more than one dimensional state spaces. Hence, the numerical computation using value iteration and policy iteration schemes is infeasible. 

 Weakly coupled MDPs and restless multi-armed bandits problems are generalizations of MDPs . They consist of multiple independent Markov processes. The decision maker is allowed to choose subset of processes with different activation levels. These  independent processes are coupled at the decision maker by budget constraints. The state of independent processes evolves whether DM acts on the process or not, according to some probability law.  Such problems fall in the class of weakly coupled MDPs or restless bandit problems. 
The objective of DM is to choose independent processes with different activation levels satisfying budget constraint at each  decision instant such that it maximizes  the long term reward function.
Obtaining a solution using value iteration and policy iteration algorithms is elusive in general.   But for special applications and examples, there are explicit closed form expressions of value functions.   
In this paper, we study and develop a framework for solving RMABs with multi-dimensional state space and multi-action arms (processes). 
 
%\subsection{Monte Carlo policy} 
A simulation based approach is popular for MDPs when value iteration and policy iterations are infeasible. Examples of such schemes are finite horizon Monte Carlo rollout policy, tree search policy algorithm and Monte Carlo tree search,  \cite{Chang05,Chang13,Kocsis2006}. The popularity of these algorithms stems from good approximation of the optimal value function. There is a trade off between computational complexity in time and state space and optimality.  %Since large state space MDPs are very difficult to solve using standard approach,  simulation based approach is another alternative approach that is extensively studied.  

In this paper we study Monte Carlo rollout policy for  MDPs and restless multi-armed bandit problem. Our work is motivated from applications in wireless communication systems, age of information (AoI), and queuing systems.

%\rahul{Need to rewrite application carefully}
Let us look at an application from communication systems. There are multiple
sources generating data, a transmitter and a set of users whose number number is
equal to that of the sources. The data from the sources is sent to the transmitter
in the form of data packets. These packets arrive at the transmitter and are
queued in a buffer with 'infinite' size. The transmitter maintains a separate queue
for each source. It transmits data to the users over wireless channels. The
number of wireless channels are less than the number of users. The transmitter
can send data only to a few of the users at any given instant. Further, the wireless
channel is Markovian, where different states represent the different channel quality
levels. Based on this there is different throughput level for each state. The queue length is
maintained by the transmitter for each of the sources. Notice that there is three
dimensional state space, i.e., queue length, channel state and arrival state. As there
are less number of channels, the objective of the transmitter is to select channels for transmission such that it maximizes the discounted infinite horizon reward (throughput) under bounded queue length. Clearly,
this is an MDP based planning problem. This is also an example of standard RMAB (two-action) when the transmitter has to decide which channel and user to select, and channel states evolve at each time instant.  The constraint at the transmitter is to select a fixed number of channels for transmission. These bandits are said to be weakly coupled by this constraint.

In the communication example, a transmitter can have choice of multiple power levels to use for transmission on a selected channel. This is applicable in the case of energy-harvesting sensor networks, \cite{Aprem13, KuChenLiu15}. For multiple users with limited channel availability and choice of multiple power level, the problem is a class of multi-action restless bandits. Here, different actions correspond to power levels available for a channel. Based on availability of battery power and it's energy constraint, a transmitter can select user channels and power levels for those channels.   

\subsection{Related work}

%\rahul{Then cite more references on RMAB, Multi-action RMAB, Optimality of RMAB,}

%\rahul{Cite more applications from age of information, hidden Markov bandit, some recent applications, security models from milind tambe}

%\rahul{Operation research application, see few papers from kaza tac submission citation}

%\rahul{cite some more refernces from queueing theory, caching and scheduling}

%\rahul{Comsnets17, CDC16, Varun journal paper, Our tac submitted, cite book from Bhatnagar}   

%\rahul{Mention about indexable and non indexable bandit}

%\rahul{Myopic policy and optimality of myopic policy}

The literature on MDP is vast and here we discuss some relevant work. The classic books on MDP are \cite{Puterman14,Bertsekas95,Bertsekas96,Kumar-Varaiya86}, they discussed value iteration, policy iteration and other variants of these algorithms in great detail for different objective functions such ass discounted reward, and average reward.  
Another approach for MDP referred to as rolling horizon procedure for MDP is studied in \cite{LeeDenardo86}. Error bounds for rolling horizon policies for discrete time MDP are derived in \cite{Lerma90} under both discounted and average reward criteria. In \cite{BertsekasTW97,BertsekasC99}, a heuristic rollout algorithm is studied for combinatorial optimization and stochastic scheduling problem.  This is a variant of the policy iteration algorithm. Rollout policies are further adapted to partially observable MDP, where states are not observed but only signals are observed and a  parallel rollout policy is introduced in \cite{Chang04,Chang13}. 

Simulation based approach using Monte-Carlo search policy is introduced in \cite{Tesauro96}. Monte Carlo simulation is performed over the tree and confidence bounds on the accuracy of stochastic policy and optimal policy are established in \cite{Kearns2002, Kearns99}. Using concentration inequalities, the number of samples required for simulation is derived. These bound scale exponentially with horizon length. This is further improved in Monte Carlo tree search algorithm, where tree search algorithm is combined with rollout horizon algorithm. There learning algorithm such as upper confidence bound scheme is employed in sampling of actions for exploration-exploitation. This UCB variant of Monte-Carlo tree search was first studied in \cite{Kocsis2006}. Also, the simulation based approach for MDP has been studied extensively for the problem of reinforcement learning, where transition and reward dynamics are unknown to decision maker, see \cite{Bertsekas96,Bertsekas20b, Sutton18}. In \cite[Chapter $5$]{Sutton18}, the Monte Carlo technique has been discussed. In reinforcement learning, the function approximation method is a  popular approach studied for large state space model. Reinforcement algorithms and rollout policies for multi-agent MDPs are studied in \cite{Bertsekas20}.

Restless multi-armed bandit and weakly coupled MDPs are generalization of MDPs, \cite{Adelman08}. RMAB is a PSPACE hard problem, \cite{Papadimitriou99} and finding optimal solution is difficult. But in \cite{Whittle88}, an index-based policy  is proposed where each bandit is assigned an index that maps state to a  real number. Using this index, $K$ arms with highest indices are played in each time step.  The popularity of this stems from near optimality of the policy~\cite{Weber90}. Recently, this work is extended to multi-action restless bandit in \cite{Glazebrook11}, where authors have introduced more that two actions for each bandits, and full indexability is defined and there are budget constraint  on actions that are allowed to activate. This is later shown to perform near optimal in \cite{Hodge15}. In all these works, the restless bandit is indexable; only then index can be computed. 
There are restless bandits which are non-indexable, this is often the case when not much  structure is imposed on the dynamics of the problem.  

Restless bandits with two-actions models are extensively studied  in applications of stochastic scheduling \cite{Nino-Mora01,Nino-Mora07,AkbarzadehMahajan20}, multi-class queueing networks \cite{Bertsimas-Nino-Mora96,Ansell03,Avrachenkov13,verloop16}, machine maintenance problem \cite{Glazebrook05, Glazebrook06}, scheduling in wireless network \cite{Borkar17b}. Restless bandit is further gernalized for hidden state MDP, i.e., partially observable MDPs. This has increased the scope of the applications to wireless opportunistic scheduling in cognitive radio \cite{LiuZhao10,LiNeely10,Meshram18,Ny08, Nino-Mora08,Nino-Mora09,Nino-Mora11}, social networks and 5G \cite{Varun18}, wireless relay networks \cite{Kaza19}, cyber physical control systems \cite{KazaLRBCPS20arxiv} , online recommendation systems \cite{Meshram17,Avrachenkov18}, and operation research \cite{Brown17}.  There are other applications emerging recently  in security \cite{QianTambe16},  age of  information \cite{Hsu18}, and health care \cite{MateTambe20arxiv, Bhattacharya18}.  
    
%\rahul{cite papers on machine replacement problem}.  
   
Though there is a lot of work on two-action restless bandits in MDP and POMDPs, there is very limited study on multi-action restless bandits \cite{Glazebrook11,Hodge15}. Further, most of the studies available, are for indexable bandits. Non-indexable bandits are very challenging and there is no general approach available. In \cite{verloop16}, the authors studied and discussed non-indexable bandits with  MDP under some assumptions on the model.  Difficulties in non-indexable bandit persist due to very less structure on the problem. This is one of the open areas for heuristic policies. In this paper we study a heuristic  Monte-Carlo rollout policy.

\subsection{Summary of the Contributions}
In this paper, we study Monte Carlo rollout policy and parallel rollout policy  for multi-dimensional MDP with multi-dimensional state space. We discuss theoretical results for rollout policies.  We next extend this study for restless multi-armed bandit problem with multi-dimensional state space and multi-action bandit model. Here, we study two classes of bandits---indexable and non-indexable bandits. For indexable bandits, we assume the existence threshold-type policy and this provides us claim indexability for each restless bandit. We use Monte-carlo rollout policy to compute an  approximate index for RMAB. Here, we consider separate cases for both two-action bandits and more than two action bandits. This is because it requires a different notion of indexability. In case of two-action bandit model, we consider standard Whittle indexability, whereas in multi-action bandit model (more than two actions), we introduce the concept of full indexability which is generalization of standard Whittle index. The concept of full indexability is inspired from \cite{Glazebrook11}. Then, a heuristic greedy approach for index computation is presented. In case of non-indexable bandit model,   we consider rollout policy with greed approach in each slot, and we do not consider any assumption on model structure. 
This is the first study of Monte-carlo rollout policies for RMAB with multi-action bandit models. We believe this can provide a reasonable   simulation based approach for complex  RMAB with least structure on the problem.

%\rahul{Organization of paper}
The rest of the paper is organized as follows. In Section~\ref{sec:prelim-MDPs}, preliminaries on MDPs are discussed for multi-dimensional state space. Simulation based Monte Carlo rollout policy and parallel rollout policy are described in Section~\ref{sec:Mote-Carlo-rollout-MDPs}. Here, we also give theoretical results. Rollout policy for restless multi-armed bandit model with multi-action and  indexable bandit is presented in Section~\ref{sec:RMAB-Rollout-indexable}, where we studied index policy and used rollout policy for index computation. Later, non-indexable restless bandit model for two action with Monte Carlo rollout policy are studied in Section~\ref{sec:RMAB-rollout-nonindexable-two-action}. This is extended  for multi-action restless non-indexable bandit in Section~\ref{sec:RMAB-rollout-nonindexable-multi-action}. Finally, we make concluding remarks and future extension in Section~\ref{sec:conclusion-remark}.

\section{Problem formulation and preliminaries for complex MDPs}
\label{sec:prelim-MDPs}

Consider a  Markov decision process, it is described by  $\mathcal{M} = \{\mathcal{S},\mathcal{A}, \mathcal{P}, \mathcal{R},\beta \}.$   The  state space $\mathcal{S} =  \prod_{i=1}^{n} S_i,$ which is $n$ dimensional and $S_i$ is subset of integer. Hence  $\mathcal{S} \in \mathbb{Z}^n.$ $\mathcal{A}$ is action space and we assume it is finite.  $\mathcal{P}$ is transition law or probability matrix which is give by $\mathcal{P} = [[P(\mathbf{y}|\mathbf{x},a)]],$ where $ \mathbf{y} = (y_1,y_2,\cdots, y_n) \in \mathcal{S},$  $\mathbf{x} = (x_1,x_2, \cdots, x_n) \in \mathcal{S},$ and $a \in \mathcal{A}.$  $\mathcal{R} = \{r(\mathbf{x},a)\}_{\{\mathbf{x} \in \mathcal{S},a \in \mathcal{A} \}}$ is immediate reward matrix.  $\beta$ is the discount parameter and  $0 < \beta <1.$ The system works in discrete time, where system state evolves at discrete time instants and action is taken at those times. Time is indexed by $t.$ Let $\mathbf{x}(t) \in \mathcal{S}$ and $a(t) \in \mathcal{A}$ denote state of system  and decision at time $t,$ respectively.  The  decision maker yields reward $r_t = r(\mathbf{x}(t),a(t))$ at time $t.$ The objective of decision maker is chose the decision to maximize long term reward function. 

%We consider  three different performance measures---discounted infinite horizon problem, undiscounted finite horizon problem and risk sensitive cost problem.   
. 
The infinite horizon discounted reward under policy $\pi$ for initial state $\mathbf{x}$ is given by 
\begin{eqnarray*}
	V_{\pi}(\mathbf{x}) := E_{\pi,\mathbf{x}} \bigg\{\sum_{t=1}^{\infty} \beta^{t-1}  r_t \bigg\}.
\end{eqnarray*}
 The policy $\pi$ is sequence of decisions, and $\pi = \{\pi(\mathbf{x}_1), \pi(\mathbf{x}_2), \cdots \}. $
Our interest is to find the  optimal policy $\pi^* \in \Pi$ such that  
\begin{eqnarray*}
	\pi^*(\mathbf{x}) \in \arg \max_{\pi \in \Pi } V_{\pi}(\mathbf{x}).   
\end{eqnarray*}

The finite horizon discounted reward under policy $\pi$ for initial state $\mathbf{x}$  is  
\begin{eqnarray*}
	V_{\pi,T}(\mathbf{x}) := E_{\pi,\mathbf{x}} \bigg\{\sum_{t=1}^{T} \beta^{t-1}  r_t \bigg\}.
\end{eqnarray*}

Our interest is to find the  optimal policy $\pi^{*}_T \in \Pi$ such that  
\begin{eqnarray*}
	\pi^{*}_T(\mathbf{x}) \in \arg \max_{\pi \in \Pi } V_{\pi,T}(\mathbf{x}).   
\end{eqnarray*}

We assume that $\vert r(\mathbf{x},a)\vert \leq B.$ Hence $V_{\pi}(\mathbf{x}) \leq \frac{B}{1-\beta}.$

\textbf{Existence of an optimal policy :} Under reasonable assumptions on state space, action space and immediate reward function, one can show the existence of optimal policy, \cite{Lerma96}. A set of such conditions include, compact state space $\mathcal{S}$ and action space $\mathcal{A}$; bounded and lower semi-continuous immediate reward function $r(\mathbf{x},a)$, and suitably defined transition probabilities.  
When state and action spaces are compact but immediate reward function is unbounded,  one can impose a notion of weighted norm and lower semi-continuity on reward function and some assumptions on transition probabilities to show the existence of optimal policy, \cite[Chapter $3,$ Section $3.3$]{Lerma96}. For our work, we proceed by assuming that all necessary conditions for existence of optimal policy hold.   

To solve the above optimization problem, the dynamic program is given as follows. 
\begin{equation*}
T V (\mathbf{x}) = \max_{ a \in A(\mathbf{x}) } \left\{r(\mathbf{x},a) + \beta E V(\mathbf{y}~|~(\mathbf{x},a)) \right\}
\end{equation*}
This is also known as Bellman operator. 
We study a stationary deterministic optimal policy. For any stationary policy $\pi \in \Pi$ we can have 
\begin{equation*}
T_{\pi} V (\mathbf{x}) =  \left\{r(\mathbf{x},\pi(a)) + \beta E V(\mathbf{y}~|~(\mathbf{x},\pi(a))) \right\}
\end{equation*}
Note that  $T$ and $T_{\pi}$ follow monotonicity properties, i.e., for any $V$ and $V^{\prime}$ such that $V(\mathbf{x})
\leq V^{\prime}(\mathbf{x})$ for all $\mathbf{x}$ any policy $\pi \in \Pi$ we have 
\begin{eqnarray*}
T V (\mathbf{x})  \leq T V^{\prime} (\mathbf{x}) \\
T_{\pi} V (\mathbf{x})  \leq T_{\pi} V^{\prime} (\mathbf{x})  
\end{eqnarray*}
Also $T$ and $T_{\pi}$ follows contraction mapping property, 
i.e., for any $V$ and $V^{\prime}$ such that $V(\mathbf{x})
\leq V^{\prime}(\mathbf{x})$ for all $\mathbf{x}$ any policy $\pi$ we have 
\begin{eqnarray*}
\max_{\mathbf{x}} \vert T V (\mathbf{x}) - T V^{\prime} (\mathbf{x})\vert \leq \beta \max_{\mathbf{x}}  \vert V (\mathbf{x}) - V^{\prime} (\mathbf{x}) \vert,  \\
\max_{\mathbf{x}} \vert T_{\pi} V (\mathbf{x}) - T_{\pi} V^{\prime} (\mathbf{x})\vert \leq \beta \max_{\mathbf{x}}  \vert V (\mathbf{x}) - V^{\prime} (\mathbf{x}) \vert.
\end{eqnarray*} 

There are two main algorithms for solving MDP---value iteration algorithm where value function is iteratively computed, by applying optimal dynamic program for each iteration and policy iteration algorithm where one starts with fixed initial policy $\pi^{n},$ evaluates the value function under fixed policy, improves that policy iteratively. 

\subsubsection{Value iteration algorithm} 
\begin{figure}[t]
	\begin{center} 		
		\begin{tikzpicture}[scale =0.65]
		\draw [->, blue, ultra thick] (1.5,2.9) --(4,2.9);
		% Policy Evaluation
		\filldraw[fill=green!40!white, draw=black, ultra thick] (4,1.7) rectangle (7,4.5);
		\node [red] at (5.5, 4) {Bellman};
		\node [red] at (5.5, 3.3) {Operation};
		\node [red] at (5.5, 2.7) {$TV_{n}$};
		\node [black] at (2.2, 3.3) {$V_{n}$};
		\node [black] at (13.5, 3.3) { If $(\vert TV_{n} - V_{n}\vert < \epsilon)$};
		\node [black] at (13.5, 2.3) { Stop: $V_{n+1}$};
		\draw [->, blue, ultra thick] (7,2.9) --(13,2.9); 
		% Rollout
		%	\draw [->, blue, ultra  thick] (12,2.9) --(12.8,2.9);
		\draw [blue, ultra thick] (10.5,2.9) --(10.5,0.3);
		\draw [ blue, ultra thick] (10.5,0.3) --(1.5,0.3);
		\draw [->, blue, ultra thick] (1.5,0.3) --(1.5,2.9);
		\node [black] at (9, 3.3) { $V_{n+1}$};
		%\node [Black] at (5, 0.6) {Feedback (ACK)};
		%	\draw [ultra thick] (4,4) --(6,4);
		%	\draw [ultra thick] (4,4) --(4,4.7);
		%	\draw [ultra thick] (6,4) --(6,4.7);
		%	\draw [ultra thick] (4,4.7) --(6,4.7);
		
		\end{tikzpicture}
	\end{center} 
	\caption{Illustration of Value iteration algorithm}
	\label{Fig.value-iter}
\end{figure}
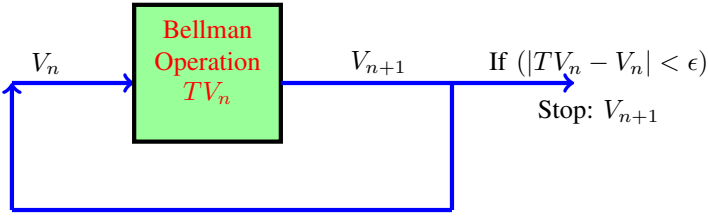

The  value iteration algorihm is given as follows. 
\begin{eqnarray}
V_{n}(\mathbf{x})= TV_{n-1}(\mathbf{x})
 = \max_{a \in A(\mathbf{x})} 
 	\left[ r(\mathbf{x}.a) + 
 	\right. \nonumber \\ \left. 
 	 \sum_{\mathbf{y} \in \mathcal{S}} p( \mathbf{y}~\vert~\mathbf{x},a) V_{n-1}(\mathbf{y})\right].
\end{eqnarray}
and next iteratively applying Bellman operation, the optimal value function is 
\begin{eqnarray*}
V^{*}(\mathbf{x}) = \lim_{n\rightarrow \infty} T^{n}V (\mathbf{x}).
\end{eqnarray*} 
The implementation of this algorithm in practice is described in Fig.~\ref{Fig.value-iter}, where value iteration algorithm stops and exit whenever $(\vert TV_{n} - V_{n}\vert < \epsilon)$ for some fixed $\epsilon > 0,$ otherwise it performs Bellman operation in loop.  

\subsubsection{Policy iteration algorithm}
We now discuss the policy iteration algorithm. In this, one starts with fixed initial base policy $\pi$ and performs policy evaluation step, computes the value function.  This is done using following iteration:
\begin{equation*}
V_{\pi}(\mathbf{x})= T_{\pi} V (\mathbf{x}) =  \left\{r(\mathbf{x},\pi(a)) + \beta E_{\pi} \left[ V(\mathbf{y}~|~(\mathbf{x},\pi(a))) \right]  \right\}
\end{equation*}
This iteration is performed till $\vert T_{\pi} V (\mathbf{x}) - V (\mathbf{x}) \vert \leq \epsilon.  $ In other word this is similar to value iteration scheme under fixed policy, i.e., value iteration under fixed policy $\pi$ at step $n$ is

\begin{equation*}
V_{n+1} (\mathbf{x}) = T_{\pi} V_{n} (\mathbf{x}) =  \left\{r(\mathbf{x},\pi(a)) + \beta E_{\pi} \left[ V_{n}(\mathbf{y}~|~(\mathbf{x},\pi(\mathbf{x}))) \right]  \right\}.
\end{equation*}

\begin{figure}
	\begin{center} 		
		\begin{tikzpicture}[scale =0.7]
		% Base policy 
		\filldraw[fill=green!40!white, draw=black, ultra thick] (0,2) rectangle (2.5,4);
		\node [black] at (1.2, 3.5) {Base}; 
		\node [black] at (1.2, 2.8) {Policy};
		\node [black] at (1.2, 2.3) {$\pi$};
		\draw [->, red, ultra thick] (2.5,2.9) --(4,2.9);
		% Policy Evaluation
		\filldraw[fill=green!40!white, draw=black, ultra thick] (4,1.7) rectangle (7,4.3);
		\node [red] at (5.5, 3.5) {Policy};
		\node [red] at (5.5, 2.8) {Evaluation};
		\node [red] at (5.5, 2.3) {$V_{\pi}$};
		% Policy improvement
		\filldraw[fill=green!40!white, draw=black, ultra thick] (8.5,1.7) rectangle (12,4.3);
		\node [red] at (10, 3.5) {Policy};
		\node [red] at (10, 2.8) {Improvement};
		\node [red] at (10, 2.3) { new policy $\widetilde{\pi}$};
		\draw [->, red, ultra thick] (7,2.9) --(8.5,2.9); 
		% Rollout
		\draw [->, blue, ultra  thick] (12,2.9) --(12.8,2.9);
		\draw [blue, ultra thick] (12.8,2.9) --(12.8,0.3);
		\draw [->, blue, ultra thick] (12.8,0.3) --(1,0.3);
		\draw [->, blue, ultra thick] (1,0.3) --(1,2);
		\node [black] at (6.5, 0.7) {Rollout policy $\widetilde{\pi}$};
		%\node [Black] at (5, 0.6) {Feedback (ACK)};
		%	\draw [ultra thick] (4,4) --(6,4);
		%	\draw [ultra thick] (4,4) --(4,4.7);
		%	\draw [ultra thick] (6,4) --(6,4.7);
		%	\draw [ultra thick] (4,4.7) --(6,4.7);
		
		\end{tikzpicture}
	\end{center} 
	\caption{Illustration of Policy iteration algorithm}
	\label{Fig.policy-iter}
\end{figure}
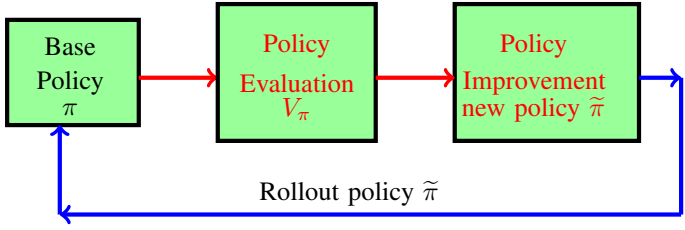

Next step is policy improvement. In preceding equation,  first part is immediate reward from feasible action $a$ and second component is value function obtained by following fixed policy $\pi.$  The new policy is found  by optimizing over actions which maximizes the immediate reward plus future value function under policy $\pi.$ This optimization is performed for all states $\mathbf{x},$ and it is given by
\begin{eqnarray*}
\widetilde{\pi}(\mathbf{x}) \in \arg\max_{a \in A(x)} \left\{r(\mathbf{x},a) + \beta E_{\pi} \left[ V_{k}(\mathbf{y}~|~(\mathbf{x},a)) \right]  \right\}
\end{eqnarray*} 
and for policy improvement  $V_{\widetilde{\pi}}(\mathbf{x}) \geq V_{\pi}(\mathbf{x})$ for all $\mathbf{x}.$
Next we repeat policy evaluation steps for policy $\widetilde{\pi},$ further this policy is improved again using improvement step. This is repeated until there is no further improvement in policy possible. 
This algorithm is illustrated in Fig.~\ref{Fig.policy-iter}. In the figure $V_{\pi}$ is the value function derived using policy $\pi.$ 

It is important to note that value iteration algorithm and policy iteration algorithm computationally expensive in practice for large state space and multi-dimensional state space. We need alternative algorithms and this is given in next section.

%\rahul{Read the following paper of Bertsekas \cite{Bertsekas20} and look at \cite[Chapter $2$]{Bertsekas20b}.}

\section{Monte Carlo rollout policy for  MDPs}
\label{sec:Mote-Carlo-rollout-MDPs}
The value iteration and policy iteration algorithms suffer from curse of dimensionality problem, because  in Bellman operation, we have to compute $E\left[ V(\mathbf{y}~|~(\mathbf{x},a))\right],$  for large multi-dimensional state-space. 
Hence we  study a simple heuristic rollout policy, where goal is to approximate $E\left[ V(\mathbf{y}~|~(\mathbf{x},a))\right]$ using Monte Carlo simulations. 

%The idea of rollout policy using Monte Carlo search for MDP is explained as follows. 

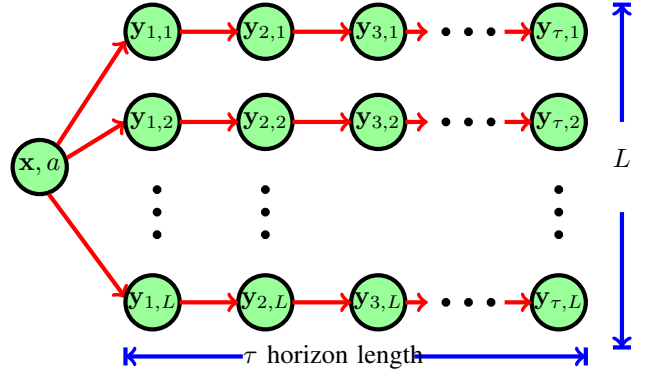
\begin{figure}
	\begin{center} 		
		\begin{tikzpicture}[scale =0.6]
		% starting state
		\filldraw[fill=green!40!white, draw=black, ultra thick] (-0.5,7) circle (0.6);
		\node [black] at (-0.5, 7) {$\mathbf{x},a$}; 
		\draw [->, red, ultra  thick] (-0.1,7.5) --(1.4,9.8);
		\draw [->, red, ultra  thick] (0.1,7.2) --(1.4,8);
		\draw [->, red, ultra  thick] (-0.3,6.4) --(1.4,4.1);
		% 1
		\filldraw[fill=green!40!white, draw=black, ultra thick] (2,10) circle (0.6); 
		\filldraw[fill=green!40!white, draw=black, ultra thick] (4.5,10) circle (0.6);
		\filldraw[fill=green!40!white, draw=black, ultra thick] (7,10) circle (0.6);
		\node [black] at (2, 10) {$\mathbf{y}_{1,1}$}; 
		\node [black] at (4.5, 10) {$\mathbf{y}_{2,1}$};
		\node [black] at (7, 10) {$\mathbf{y}_{3,1}$};
		\filldraw[fill=black, draw=black, ultra thick] (8.5,10) circle (0.05);
		\filldraw[fill=black, draw=black, ultra thick] (9,10) circle (0.05);
		\filldraw[fill=black, draw=black, ultra thick] (9.5,10) circle (0.05);
		\filldraw[fill=green!40!white, draw=black, ultra thick] (11,10) circle (0.6);
		\node [black] at (11, 10) {$\mathbf{y}_{\tau,1}$};
		\draw [->, red, ultra  thick] (2.6,10) --(3.9,10);
		\draw [->, red, ultra  thick] (5.1,10) --(6.4,10);
		\draw [->, red, ultra  thick] (7.6,10) --(8.1,10);
		\draw [->, red, ultra  thick] (9.8,10) --(10.4,10);
		% 2 
		\filldraw[fill=green!40!white, draw=black, ultra thick] (2,8) circle (0.6);
		\filldraw[fill=green!40!white, draw=black, ultra thick] (4.5,8) circle (0.6);
		\filldraw[fill=green!40!white, draw=black, ultra thick] (7,8) circle (0.6);
		\filldraw[fill=black, draw=black, ultra thick] (8.5,8) circle (0.05);
		\filldraw[fill=black, draw=black, ultra thick] (9,8) circle (0.05);
		\filldraw[fill=black, draw=black, ultra thick] (9.5,8) circle (0.05);
		\filldraw[fill=green!40!white, draw=black, ultra thick] (11,8) circle (0.6);
		\node [black] at (2, 8) {$\mathbf{y}_{1,2}$}; 
		\node [black] at (4.5, 8) {$\mathbf{y}_{2,2}$};
		\node [black] at (7, 8) {$\mathbf{y}_{3,2}$}; 
		\node [black] at (11, 8) {$\mathbf{y}_{\tau,2}$};
		\filldraw[fill=black, draw=black, ultra thick] (2.1,5.5) circle (0.05);
		\filldraw[fill=black, draw=black, ultra thick] (2.1,6) circle (0.05);
		\filldraw[fill=black, draw=black, ultra thick] (2.1,6.5) circle (0.05);
		\filldraw[fill=black, draw=black, ultra thick] (4.5,5.5) circle (0.05);
		\filldraw[fill=black, draw=black, ultra thick] (4.5,6) circle (0.05);
		\filldraw[fill=black, draw=black, ultra thick] (4.5,6.5) circle (0.05);
		\filldraw[fill=black, draw=black, ultra thick] (11,5.5) circle (0.05);
		\filldraw[fill=black, draw=black, ultra thick] (11,6) circle (0.05);
		\filldraw[fill=black, draw=black, ultra thick] (11,6.5) circle (0.05);
		\draw [->, red, ultra  thick] (2.6,8) --(3.9,8);
		\draw [->, red, ultra  thick] (5.1,8) --(6.4,8);
		\draw [->, red, ultra  thick] (7.6,8) --(8.1,8);
		\draw [->, red, ultra  thick] (9.8,8) --(10.4,8);
		% L
		\filldraw[fill=green!40!white, draw=black, ultra thick] (2,4) circle (0.6);
		\filldraw[fill=green!40!white, draw=black, ultra thick] (4.5,4) circle (0.6);
		\filldraw[fill=green!40!white, draw=black, ultra thick] (7,4) circle (0.6);
		\filldraw[fill=black, draw=black, ultra thick] (8.5,4) circle (0.05);
		\filldraw[fill=black, draw=black, ultra thick] (9,4) circle (0.05);
		\filldraw[fill=black, draw=black, ultra thick] (9.5,4) circle (0.05);
		\filldraw[fill=green!40!white, draw=black, ultra thick] (11,4) circle (0.6); 
		\node [black] at (2, 4) {$\mathbf{y}_{1,L}$}; 
		\node [black] at (4.5,4) {$\mathbf{y}_{2,L}$};
		\node [black] at (7, 4) {$\mathbf{y}_{3,L}$}; 
		\node [black] at (11, 4) {$\mathbf{y}_{\tau,L}$};
		\draw [->, red, ultra  thick] (2.6,4) --(3.9,4);
		\draw [->, red, ultra  thick] (5.1,4) --(6.4,4);
		\draw [->, red, ultra  thick] (7.6,4) --(8.1,4);
		\draw [->, red, ultra  thick] (9.8,4) --(10.4,4);
		\draw [blue, ultra  thick] (11.6,3.0) --(11.6,2.6);		
		\draw [->, blue, ultra  thick] (7.8,2.8) --(11.6,2.8);
		\node [black] at (6, 2.8) {$\tau$ horizon length};
		\draw [blue, ultra  thick] (1.4,3.0) --(1.4,2.6);		
		\draw [->, blue, ultra  thick] (4,2.8) --(1.4,2.8);
		\draw [->,blue, ultra  thick] (12.4,8) --(12.4,10.6);
		\draw [->,blue, ultra  thick] (12.4,6) --(12.4,3);		
		\draw [blue, ultra  thick] (12.2,10.6) --(12.6,10.6);
		\draw [blue, ultra  thick] (12.2,3) --(12.6,3);
		\node [black] at (12.4, 7.2) {$L$};
	%	\node [black] at (13, 7.2) {$L$ sampled};
	%	\node [black] at (13, 6.7) {trajectories};
	%	\node [black] at (13, 6.2) {base policy $\mu$};
		\end{tikzpicture}
	\end{center} 
	\caption{Illustration of Monte Carlo Rollout policy}
	\label{Fig.rollout-policy}
\end{figure}

%This is a simulation based approach.
We start with state-action $(\mathbf{x},a),$ and generate  trajectories using a base policy $\pi.$ We assume that  $L$ number of  trajectories are generated. These trajectories are obtained for depth of horizon length $\tau.$   Each trajectory consists of  evolution of states, observations of rewards using on policy $\pi.$ Here, next state can be obtained from transition model or generative model under policy $\pi.$   We compute the discounted reward accrued from a trajectory over $\tau$ horizon, further this yields empirical discounted reward from policy $\pi$ over $L$ trajectories. We approximate this empirical reward with $E V(\mathbf{x}+1~|~(\mathbf{x},a)).$ The optimal action for given state  $\mathbf{x}$ is chosen based on immediate reward and empirical discounted reward obtained using Monte-Carlo rollout policy. The policy $\pi$ is updated to new policy $\widetilde{\pi}$ based on optimal action choose for $\mathbf{x}.$ This is described in Fig.~\ref{Fig.rollout-policy}. 

%In each stage, the empirical discounted reward obtained from following policy $\pi$ is backed up.  In this model, next state from given state and action according to state transition dynamic $P(y~|~x,a).$ This is  generative model approach, where next state transition and rewards are generated using model. Moreover this reduces computation.   

Mathematical detail is as follows. Let $Q^{l}_{\pi,t}(\mathbf{x},a,\mathbf{y}_{t,l},\pi(\mathbf{y}_{t,l}))$  be the state action value function for $l$th sample at time step $t$ under policy $\pi$ when initial state and action is $(\mathbf{x},a).$ $\mathbf{y}_{t,l}$ denotes the 
 state at time step $t$ for $l$th sample. The action under policy $\pi$ for that state is $\pi(\mathbf{y}_{t,l}).$
  Hence $Q^{l}_{\pi,t}(\mathbf{x},a,\mathbf{y}_{t,l},\pi(\mathbf{y}_{t,l})) = r(\mathbf{y}_{t,l}, \pi(\mathbf{y}_{t,l})). $
  
 When $\tau=1,$ this  rollout MC policy is a simple approximation to one-step lookahead policy and it given as follows. 
\begin{eqnarray*}
	\widetilde{Q}_{\pi,1}(\mathbf{x},a) &=&r(\mathbf{x},a) + \beta\left[    \frac{1}{L} \sum_{l=1}^{L}   Q_{\pi,1}^{l}(\mathbf{x},a,\mathbf{y}_{1,l}, \pi(\mathbf{y}_{1,l})) \right], \\
	\widetilde{\pi}(\mathbf{x}) &\in& \arg \max_{a \in A} \widetilde{Q}_{\pi,1}(\mathbf{x},a).
\end{eqnarray*}

For $\tau >1,$ total discounted reward for $l$th trajectory running upto horizon length $\tau$ with policy $\pi$ is 
\begin{eqnarray*}
	\overline{Q}_{\pi, \tau}^{l}(\mathbf{x},a) 
	= \sum_{t=1}^{\tau} \beta^{t-1} Q_{\pi,t}^{l}(\mathbf{x},a,\mathbf{y}_{t,l}, \pi(\mathbf{y}_{t,l})).
\end{eqnarray*}
The state action value function estimate from this policy is 
\begin{eqnarray*}
	\widetilde{Q}_{\pi,\tau}(\mathbf{x},a) &=& r(\mathbf{x},a) + \beta\left[    \frac{1}{L} \sum_{l=1}^{L}    \overline{Q}_{\pi, \tau}^{l}(\mathbf{x},a) \right],
\end{eqnarray*}
and new policy is 
\begin{eqnarray*}
	\widetilde{\pi}(\mathbf{x}) \in \arg \max_{a \in \overline{A}} \widetilde{Q}_{\pi,\tau}(\mathbf{x},a).
\end{eqnarray*}

\begin{remark}
	\begin{itemize} 
		\item This is an online rollout policy algorithm. It is very important to select good policy $\pi.$ Instead of fixed policy $\pi,$ one can use randomized policy, where actions (arms) are picked at given state randomly according to some fixed distribution. 
		\item If the decision maker has computational budget constraint and  bounded rational with limited memory, then online rollout policy iteration is better choice for large state space model. 
		\item Though this is a simple algorithm, this algorithm can be far off from the optimal and in later subsection we characterize this suboptimality gap and provide some bounds. 
		\item Note that rollout policy algorithm is a variant of policy iteration algorithm.
		% \rahul{Need to explain more carefully.}
	\end{itemize}
\end{remark}

\subsection{Monte Carlo parallel rollout policy for MDPs} 
We now consider another variant of rollout policy algorithm.  In practice, it is challenging to figure out good policy $\pi.$ This motivates to look for alternatives. One such alternative is parallel rollout policy which make use of addtional parallel computational architecture. In this,  different policies are evaluated using parallel architecture starting from given state-action and value estimate are obtained by running Monte Carlo search for $L$ trajectories upto $\tau$ horizons.  Recall that total discounted reward for $l$th trajectory running upto horizon length $\tau$ with policy $\pi$ is 
\begin{eqnarray*}
	\overline{Q}_{\pi, \tau}^{l}(\mathbf{x},a) 
	= \sum_{t=1}^{\tau} \beta^{t-1} Q_{\pi,t}^{l}(\mathbf{x},a,\mathbf{y}_{t,l}, \pi(\mathbf{y}_{t,l})).
\end{eqnarray*}
The taking empirical averaging of this, we get  
\begin{eqnarray*}
	\widehat{Q}_{\pi,\tau}(\mathbf{x},a) = \frac{1}{L}\sum_{l=1}^{L} \overline{Q}_{\pi, \tau}^{l}(\mathbf{x},a).
\end{eqnarray*}
This can be evaluated for finite number of policies $\pi \in \Lambda \subset \Pi,$ we assume that size of set $\Lambda$ is finite. Then  state-action value function using following procedure obtained and the optimal policy for given state is found.
\begin{eqnarray*}
	\widetilde{Q}_{\pi,\tau,pr}(\mathbf{x},a) &=&  r(\mathbf{x},a) + \beta \max_{\pi \in \Lambda} \widehat{Q}_{\pi,\tau}(\mathbf{x},a) \\
	\pi_{pr}^{\prime}(\mathbf{x}) & \in& \arg \max_{a \in A} \widetilde{Q}_{\pi,\tau,pr}(\mathbf{x},a).
\end{eqnarray*}
\begin{remark}
	\begin{itemize}
		\item Parallel rollout policy is well suited when decision maker does not have a good policy. With additional cost of parallel architecture one may able to provide good approximation to optimal value estimate.
		\item We observe that in order to obtain optimal value estimate and determine the optimal actions, while running rollout policy algorithm mentioned above, the decision maker need to have some knowledge about good policy. This knowledge may not be available at decision maker, and then this will lead to starting with some uniform random fixed policy. This may result into  suboptimal outcome.
		
		% Is there other than parallel rollout policy to find out good policy? This further motivate us to study Monte-Carlo Tree Search algorithm. This is explained section. 
		\item In vanilla Monte Carlo rollout policy, actions are played according to fixed policy for given state. While selecting base policy $\pi,$ we can exploit the structure of the problem. 
		
		\item Another variant is to consider a stochastic policy instead of fixed policy $\pi$ where actions for given state is played according to probability distribution. Note that the for this case stochastic algorithm may give better advantage and lead to near optimal solution but it requires large horizon length $\tau$ and running time to be exponential in horizon length, \cite[Theorem $1$]{Kearns2002}. The algorithm is called as sparse sampling algorithm. 
		
		%We also consider another variant of these policies are stochastic policy
	\end{itemize} 
\end{remark}

%\rahul{It would be nice to have some picture here for rollout policy}

%\rahul{Application of this for multi-dimensional state space problem, or to risk sensitive cost problem.}

\subsection{Results on Monte Carlo rollout policies}
\label{sec:rollout-MDP}
In this section, we present  theoretical guarantees for Monte carlo rollout policies.  
The objective is to measure the degree of suboptimality of Monte Carlo rollout policies for finite horizon $\tau $ with respect to optimal policy. 

We first describe results from \cite[Chapter $6$, sectioon $6.2.2$]{Bertsekas96}, which gives bound on  approximation of optimal value function and value function under fixed stationary policy $\pi.$ This is worst case bound.  This bound is improved in \cite{Singh94}, where the difference between optimal value function and one step greedy policy with approximate value function is dependent on discount parameter $\beta.$ These provide insight for  Monte Carlo rollout policies. 

Consider an approximate policy iteration, in that the sequence of stationary policy $\pi_n$ is generated and corresponding approximate value function $V_n$ which satisfies 
\begin{eqnarray*}
	\max_{\mathbf{x}} \vert V_n(\mathbf{x}) - V_{\pi^n}(\mathbf{x}) \vert &\leq& \epsilon \\
	\max_{\mathbf{x}} \vert (T_{\pi^{n+1}} V_{n})(\mathbf{x}) - (TV_n)(\mathbf{x})  \vert  &\leq& \delta. 
\end{eqnarray*}

$T_{\pi^{n+1}}$ is one step evaluation under policy $\pi^{n+1}.$ Define $V^*$ as the optimal value function. From \cite[Chapter $6$, sectioon $6.2.2$]{Bertsekas96}, the worst case bound on approximate policy iteration is given in following proposition.
	\begin{proposition}
		\begin{eqnarray*}
			\lim\sup_{n\rightarrow \infty} \max_{\mathbf{x}} \vert V_{\pi^n}(\mathbf{x}) - V^*(\mathbf{x}) \vert  
			\leq \frac{\delta +2 \beta \epsilon }{(1-\beta^2)}.
		\end{eqnarray*}
	\end{proposition}
 This is scaling with $\frac{1}{(1-\beta^2)},$	and for small values of discount parameter $\beta,$ this can be very bad.   Thus this is a pessimistic bound.
 
This result is further improved in \cite{Singh94}. Assume that  approximate value function $\widetilde{V}$ is $\epsilon$ close to optimal value function $V^*$ Then greedy policy for $\widetilde{V}$ has good approximation to optimal policy. This is given in following proposition. 
 
 \begin{proposition}[Improved bound \cite{Singh94}]
 	If approximate $\widetilde{V}$ is such that for all $\mathbf{x} \in \mathcal{S},$ $\vert V^*(\mathbf{x}) - \widetilde{V}(\mathbf{x})\vert \leq \epsilon,$ and $\pi_{\widetilde{V}}$ is greedy policy for.  $\widetilde{V},$ i.e.,
 	\begin{eqnarray*}
 		\pi_{\widetilde{V}}(\mathbf{x}) = \arg\max_{a \in \mathcal{A}} \left[ r(\mathbf{x},a) + \beta E[\widetilde{V}(y)~|~(\mathbf{x},a)]  \right] 
 	\end{eqnarray*}
 	then for all $\mathbf{x} \in \mathcal{S}$
 	\begin{eqnarray*}
 		\vert V^*(\mathbf{x}) - \widetilde{V}_{\pi_{\widetilde{V}}}(x) \vert \leq \frac{2 \beta \epsilon }{1- \beta}.
 	\end{eqnarray*}
 \label{prop:improved-bound-approx-val}
 \end{proposition}
 Here the immediate reward from state-action is assumed to be known. This is not known, but bound is known, then using this there can be obtain bound on approximation of value function.  
 \begin{corollary}
 	For $\vert r(\mathbf{x},a)-  \widetilde{r}(\mathbf{x},a) \vert  \leq \alpha.$ Then  
 	$ \vert 	V^*(\mathbf{x}) - \widetilde{V}_{\pi_{\widetilde{V}}}(x) \vert  \leq \frac{2 \beta  \epsilon  +2 \alpha}{1- \beta}$
 	Here $\widetilde{r}$ is approximate reward from given state-action.
 \end{corollary}

\begin{remark}
	If approximate value function $\widetilde{V}$ for given state-action is computed using Monte-Carlo rollout policy, then we  can have greedy policy approximation to optimal value function $V^*$ from Proposition~\ref{prop:improved-bound-approx-val}.
\end{remark}
 
In the preceding approximation, we have not specified number of time horizon needed to get fixed level of accuracy $\epsilon$ under fixed policy $\pi.$   This approximation will be given next few results. 
 $V_{\pi,\tau}$ is finite horizon discounted value function under policy $\pi$ starting from state $\mathbf{x}$ and this is given by 
\begin{eqnarray*}
	V_{\pi,\tau}(\mathbf{x}) = \mathrm{E}\left[ \sum_{t=0}^{\tau-1} \beta^{t}  \left( r(\mathbf{x}(t),a^{\pi}(t)) \right) \right].
\end{eqnarray*}

Following result is adapted from \cite{Lerma90}. Result provides bound on optimal value function and finite horizon value function under policy $\pi.$
\begin{lemma}
	\begin{eqnarray*}
		0 \leq V^*(\mathbf{x}) - V_{\pi,\tau}(\mathbf{x}) \leq \frac{R_{\max}}{1-\beta} \beta^{\tau}.
	\end{eqnarray*}
%	Intuitively this provides a degree of suboptimality of rolling horizon policy.
\label{lemma:subopt-rollout1}
\end{lemma}
Proof of this is straightforward, but clarity purpose proof  is given in  Appendix~\ref{app:lemma-subopt-rollout1}.
$V^{*}_{\tau}$ is the optimal discounted reward for finite horizon $\tau$ and  
$V^{*}_{\tau}(\mathbf{x}) = \sup_{\pi \in \Pi} V_{\pi,\tau}(\mathbf{x}).$ Let $B(\mathcal{S})$ be the set of all value functions.  In the next result, we provide bound on difference of value function under policy $\pi$ and an optimal value function for infinite horizon. 
%\end{eqnarray}
\begin{lemma}
		Given value function $V \in B(\mathcal{S})$ such that for function
	 $|V_{\tau-1}^*(\mathbf{x}) - V(\mathbf{x})| \leq \epsilon $ for all $\mathbf{x} \in \mathcal{S}$ and consider a policy $\pi \in \Pi$ such that $T_{\pi}(V) = T(V).$  	
	\begin{eqnarray*}
		0 \leq V^*(\mathbf{x}) - V_{\pi}(\mathbf{x}) \leq \frac{R_{\max}}{(1-\beta)} \beta^{\tau} + \frac{2 \beta \epsilon }{1-\epsilon}.
	\end{eqnarray*}
	\label{lemma:error-approx-policy-pi}
\end{lemma}

This result is derived in \cite[Theorem $5.1$]{Chang13}.  For sake of clarity, we provide proof in Appendix~\ref{app:lemma-error-approx-policy-pi}.
	
%Next going from finite horizon approximation to infinite horizon approximation under any policy $\pi,$ and the suboptimality of policy $\pi$ is given in next Lemma. We first define finite horizon discounted reward under policy $\pi$ and this is given byfor $x_i(0) =x_i$ and $x = (x_1,\cdots, x_N).$ The optimal discounted reward for finite horizon $H$ is 

\begin{remark}
\begin{itemize}
  \item For finite horizon, the distance between optimal value function and any value function is bounded by term $\epsilon.$ Then for any policy $\pi$ we say that infinite horizon optimal value function and the value function under policy $\pi$ can be suitably bounded. Moreover for this bound as horizon length sufficiently increases, the bounds are going to be $\frac{2 \beta \epsilon}{ 1- \epsilon}.$ This indicates the suboptimality of value function under policy $\pi.$
  \item Note that this bound is tight if discount parameter $\beta$ is away from $1,$   $\epsilon$ is small and length of horizon, i.e., $\tau$ is large.
  \item The bound in Lemma~\ref{lemma:error-approx-policy-pi} is useful while coming up with bound on rollout policy. 
\end{itemize}
\end{remark}

%\rahul{Comment with respect to number of arms in restless mab problem. If possible also mention connection to POMDP. It would be nice to provide some directions.} 

\begin{corollary}
For every $\epsilon > 0,$ there  exists $\tau >1$ such that 
$V_{\pi}(\mathbf{x})  - V_{\pi_{ro,\tau}} (\mathbf{x}) \leq  \epsilon.$ 
\end{corollary}
This is a standard consequence of expansion of value iteration using policy $\pi$ and rollout policy with finite horizon, $\pi_{ro,\tau}.$ This suggests that for sufficiently long horizon, one can approximate the value function using rollout policy.  

More importantly in next lemma, we can specify the rolling horizon length for given $\epsilon, \beta$ and $R_{\max}.$  
\begin{lemma}
	For any $\epsilon > 0$ if $\tau > 1+ \log_{\beta}\frac{\epsilon (1-\beta)}{R_{\max}}$ then for all $x \in X$
	\begin{eqnarray*}
		V_{\pi_{ro,\tau}} (\mathbf{x}) \geq V_{\pi}(
		\mathbf{x}) -\epsilon.
	\end{eqnarray*}
	\label{lemma:large-horizon-rollout-pi}
\end{lemma}
The proof is straightforward, it is given in  Appendix~\ref{app:lemma-large-horizon-rollout-pi}. 
\begin{corollary}
	Using Lemma \ref{lemma:error-approx-policy-pi} and \ref{lemma:large-horizon-rollout-pi}, we can measure the suboptimality of finite horizon rollout policy. If $\sup_{\mathbf{x} \in \mathcal{S}} \vert V^{*}(\mathbf{x}) - v_{\pi,H-1}(\mathbf{x}) \vert < \epsilon$ then we can have 
	\begin{eqnarray*}
		0 \leq V^*(\mathbf{x}) - V_{\pi_{ro},H}(\mathbf{x}) \leq \beta^{H} \frac{R_{\max}}{1-\beta} + \frac{2\beta \epsilon}{1-\beta}.   
	\end{eqnarray*}
\end{corollary}

%\rahul{Comment on bound for RMAB problem and POMDP.}

In the rollout policy, we have used base policy which is fixed. Often knowing good based policy is very difficult. 
%Can one do better than rollout policy when there is no knowledge of good base policy? 
We consider an alternative approach and one such method is parallel rollout policy. With additional computational cost, we can improve on rollout policy. Recall that in parallel rollout policy, multiple based policy are simulated and best among them is selected.

Thus 
\begin{eqnarray*} 
  \pi_{pr,\tau}(\mathbf{x}) \in \max_{a \in \mathcal{A}}  \mathrm{E}\left[ r(\mathbf{x},a) + \beta \max_{\pi \in \Pi} V_{\pi,\tau-1}(\mathbf{y})  \right].
\end{eqnarray*}

\begin{corollary}
	For any $\epsilon > 0$ there exists $\tau$ such that 
	$\max_{\pi \in \Pi} V_{\pi}(\mathbf{x}) - V_{pr,\tau}(\mathbf{x}) < \epsilon$ for $\mathbf{x} \in \mathcal{S}.$
\end{corollary}

\begin{lemma}
	For $\pi_{pr,\tau}$ defined on non empty finite subset $\Lambda \in \Pi,$ given  any $\epsilon > 0$ if 
	\begin{eqnarray*}
		\tau > 1+ \log_{\beta}\frac{\epsilon (1-\beta)}{R_{\max}},
	\end{eqnarray*}
	then for all $\mathbf{x} \in \mathcal{S}$
	\begin{eqnarray*}
		V_{\pi_{pr,H}} (\mathbf{x}) \geq \max_{\pi \in \Lambda} V_{\pi}(\mathbf{x}) -\epsilon.
	\end{eqnarray*}
	
\end{lemma}

Parallel rollout policy   studied in  \cite{Chang13} and present results from it.

\begin{property}
	If 
	\begin{eqnarray*}
		\sup_{\mathbf{x} \in \mathcal{S}} \vert V^{*}(\mathbf{x}) - \max_{\pi \in \Lambda} V_{\pi,\tau-1}(\mathbf{x}) \vert \leq \epsilon
	\end{eqnarray*}
	then 
	\begin{eqnarray*}
		0 \leq V^*(\mathbf{x}) - V_{\pi_{pr},\tau}(\mathbf{x}) \leq \beta^{\tau} \frac{R_{\max}}{1-\beta} + \frac{2\beta \epsilon}{1-\beta}.   
	\end{eqnarray*}
	for all $\mathbf{x} \in \mathcal{S}.$
\end{property}

\begin{lemma}
	For given non-empty set $\Lambda \subset   \Pi$ for parallel rollout policy 
	\begin{eqnarray*}
		V_{pr}(\mathbf{x})  \geq \max_{\pi \in \Lambda} V_{\pi} (\mathbf{x}),  &  & \mbox{$ \mathbf{x} \in \mathcal{S}.$} 
	\end{eqnarray*}
\end{lemma}

\begin{remark}
	Note that the degree of suboptimality depends on the discounter parameter $\beta,$ if  $\beta$ is away from $1$ then approximation can be good. Using parallel architecture, one can make efficient utilization of computational power.  	
\end{remark}

%\section{Trajectory Tree Method} 

%\rahul{Comment on restless multi-armed bandit and POMDP.} 
%Idea of rollout policy can be extended to POMDP and RMAB. 
%\rahul{How to use rollout policies when there is structure in the problem or when there exists a threshold policy behavior?}

%\rahul{What is new theory or innovation can be introduced in simple rollout policy of parallel rollout policy?}

\section{Restless multi-armed bandits and Monte Carlo rollout policy}
\label{sec:RMAB-Rollout-indexable}
We now consider problem of restless multi-armed  bandit and study various heuristic policies. 
Suppose  RMAB has $N$ independent arms and RMAB is described by  $\mathcal{M}_B = \{\mathcal{S},\mathcal{A}, \mathcal{P}, \mathcal{R}, N, \beta\}.$ 
 
The  state space of bandit $\mathcal{S} =  \prod_{i=1}^{N} \mathcal{S}_i,$  where $\mathcal{S}_i$ represents  state of arm $i,$ this can be further $n$ dimensional, i.e., $\mathcal{S}_i = \prod_{j=1}^{n}S_{ij}$ and $S_{ij}$ is subset of integer. Thus for each arm $i,$  $\mathcal{S}_i \subset \mathbb{Z}^n.$ The action space   $\mathcal{A} =  \prod_{i=1}^{N} \mathcal{A}_i,$ here $\mathcal{A}_i$ is the action of  arm $i.$ The transition  probabiltiy matrix is given by $\mathcal{P} = \{P_i(\mathbf{y}_i~|~\mathbf{x}_i,a_i)\}_{i=1}^{N},$   and $\mathbf{x}_i = (x_{i,1}, \cdots, x_{i,n} ),$ $\mathbf{y}_i = (y_{i,1}, \cdots, y_{i,n}),$ $\mathbf{x}_i, \mathbf{y}_i \in \mathcal{S}_i.$ and $a_i \in \mathcal{A}_i.$
We write $P(Y~|X,\mathbf{a}) = \prod_{i=1}^{N} P_i(\mathbf{y}_i~|~\mathbf{x}_i,a_i),$ $X = \{\mathbf{x}_1, \mathbf{x}_2, \cdots,\mathbf{x}_N\},$ $Y =\{\mathbf{y}_1,\mathbf{y}_2,\cdots,\mathbf{y}_N\},$ and $\mathbf{a} =\{a_1,a_2,\cdots, a_N\}.$ The immediate reward is $\mathcal{R} = \{ r_i(\mathbf{x}_i,a_i)\}_{i=1}^{N} \in \mathbb{R}^N.$ 

The action space of arm $i,$ $\mathcal{A}_i$ is assumed to be discrete and finite, size of set $\mathcal{A}_i$ is $m < \infty.$ This represent the class of multi-action restless bandit problem. For case $m=2,$ we can have $\mathcal{A}_i = \{0,1\},$ where $a_i = 0$ correspond to not playing of arm $i$ and $a_i = 1$ correspond to  playing of arm $i.$ This is the most frequently studied model for RMAB. We here study generalized RMAB, and allow more that $2$ actions for each arms. Multiple-actions describe the activity level, $a_i \in \mathcal{A}_i$ represents the activity of arm $i$ selected by decision maker. There are finite activity levels. $\beta$ is the discount parameter, $0<\beta < 1.$ 
We suppose that time is slotted and it is indexed by $t.$ 
Let $a_i(t) \in \mathcal{A}_i$ denote the action (activity level) for arm $i$ selected by decision maker.
For case of $m=2,$  $a_i(t) =1$ when arm $i$ is played in slot $t$ and $a_i(t) =0$ when arm is not played in slot $t.$ 
$\mathbf{x}_i(t)$ denotes the state of arm $i$ at the beginning of slot $t.$ The immediate reward  accrued to decision maker from action $a_i(t)$ is $r_i(\mathbf{x}_i(t),a_i(t)).$ 
Let $\pi = \{ \pi(t) \}_{t\geq 1}$ be the policy  that maps history to actions for each arm and it is defined as follows: $\pi(t): H_t \rightarrow  \mathcal{A}.$
Here,  $H_t$ denotes the history of state, actions and observed rewards up to time slot $t,$ i.e., $\{\mathbf{x}_i(s), a_i(s), r_i(\mathbf{x}_i(s), a_i(s))\}_{1 \leq i \leq N, 1 \leq s <t}.$ 
The total expected discounted infinite horizon reward 
under policy $\pi$ with initial state $X = (\mathbf{x}_1, \cdots, \mathbf{x}_N)$ is given by 
\begin{eqnarray*}
	V_{\pi}(X) := \mathrm{E}_{\pi,X} \bigg\{\sum_{t=1}^{\infty} \beta^{t-1} \left( \sum_{i=1}^{N} r_i(\mathbf{x}_i(t), a_i(t))\right)\bigg\}.
\end{eqnarray*}
There is total budget constraint (activity constrant) at the decision maker for each time slot $t$ and this is given by 
\begin{eqnarray*}
\sum_{i=1}^{N}a_i(t) \leq K,  \ \ \ \  a_i(t) \in \mathcal{A}_i.
\end{eqnarray*} 

%There are allowed to play $K$ arms in any given time slot, $1\leq K < N.$ Thus,  $\sum_{i=1}^{N}a_i(t) = K.$ The state of arm $i$ at time slot $t$ is denoted by $x_i(t).$ Let $\pi = \{ \pi(t) \}_{t\geq 1}$ be the policy  that maps history to actions and it is defined as follows: $\pi(t): H_t \rightarrow \mathcal{K}$ such that $|\mathcal{K}| = K,$  and $ \mathcal{K} \subset \mathcal{N} = \{1,2,\cdots,N\}.$  The infinite horizon discounted reward under policy $\pi$ for initial state $x$ is given by 

Then objective is to find the optimal  policy  $\pi^*:$ 
\begin{eqnarray*}
	\pi^* &=& \arg \max_{\pi} V_{\pi}(X).   \\
	&\text{ s.t. } &\sum_{i=1}^{N}a_i(t) \leq K,  \ \ \ \  a_i(t) \in \mathcal{A}_i.
\end{eqnarray*}
for all $X \in \mathcal{S}.$ 
$V_{\pi}: \mathcal{S} \rightarrow \mathbb{R}$ is the value function under policy $\pi$. The optimal value function $V(X)$ is 
\begin{eqnarray*}
	V(X) &=& \max_\pi V_{\pi}(X) \\  
	&\text{ s.t. } &\sum_{i=1}^{N}a_i(t) \leq K,  \ \ \ \  a_i(t) \in \mathcal{A}_i.
\end{eqnarray*}

%We now comment on existence of optimal policy. Under reasonable assumption on state space, action space and immediate reward function, one can show the existence of optimal policy, \cite{Lerma96}. Hence we assume that state space and action space is compact, immediate reward function is continuous in state and action. When state space, action space is not compact, and immediate reward function is unbounded, then one can introduce notion of weighted norm concept and lower semi continuity on reward function and some assumptions on transition probabilities to show the existence of optimal policy, \cite[Chapter $3$]{Lerma96}.
 We assume that all necessary conditions hold for existence of optimal policy.   
Then the dynamic program that solves optimal value function   is given by 
\begin{eqnarray*}
	V(X)&=& \max_{\mathbf{a} \in \overline{A}} \left[ \sum_{i=1}^{N} r_i(\mathcal{x}_i, a_i) + \beta \sum_{Y \in \mathcal{S}} P(Y~|X,\mathbf{a}) V(Y) \right].
\end{eqnarray*}
\begin{eqnarray*}
	V(X)&=& \max_{\mathbf{a} \in \overline{A}} \left[ \sum_{i=1}^{N} r_i(\mathbf{x}_i, a_i) +  \right. \\ & & \left.
	\beta \sum_{Y=(\mathbf{y}_1,\cdots, \mathbf{y}_N) \in \mathcal{S}} \prod_{i=1}^{N} P_i(\mathbf{y}_i~|~\mathbf{x}_i,a_i) V(Y) \right]. 
\end{eqnarray*}
Here, $\overline{A} = \{\mathbf{a} = (a_1, \cdots, a_N) \in \mathcal{A} : \sum_{i=1}^{N}a_i \leq K \}.$
The value iteration algorithm can be written as follows. 
\begin{eqnarray*}
	V_{t+1}(X)&=&  \max_{\mathbf{a} \in \overline{A}} \left[ \sum_{i=1}^{N} r_i(x_i, a_i) + \beta 
	\right. \\ & & \left. 
	\sum_{Y=(\mathbf{y}_1,\cdots, \mathbf{y}_N) \in \mathcal{S}} \prod_{i=1}^{N} P_i(\mathbf{y}_i~|~\mathbf{x}_i,a_i) V_t(Y) \right]. 
\end{eqnarray*}

In the policy iteration algorithm, there are two steps, 1) policy evaluation and 2) policy improvment.  In the policy evaluation step, initial fixed policy $\pi$ is runned, using this policy the following value function computation is performed. 
\begin{eqnarray*}
	V_{\pi}(X,a) &=& \left[ \sum_{i=1}^{N} r_i(\mathbf{x}_i, a_i) + \beta 
\right. \\ &	& \left.    \sum_{Y=(\mathbf{y}_1,\cdots, \mathbf{y}_N) \in \mathcal{S}} \prod_{i=1}^{N} P_i(\mathbf{y}_i~|~\mathbf{x}_i,\pi(X)) V_{\pi}(Y) \right]. 
\end{eqnarray*}
Next step is policy improvement, where policy is improved 
\begin{eqnarray*}
	\pi^{\prime}(X) \in \arg \max_{\mathbf{a} \in \mathcal{A}} V_{\pi}(X,a), & &  \& \ \ V_{\pi^{\prime}}(X) \geq  V_{\pi}(X) 
\end{eqnarray*}
for all $X \in \mathcal{S}.$ Then one updates the old policy with new one and repeats policy evaluation and improvement steps until no more further improvements are possible.

\begin{remark}
	
	\begin{itemize} 
		\item Note that value iteration and policy iteration in the preceding problem have the following difficulties: 1) state space of each arm is large, and hence it is computationally difficult to run value iteration or policy iteration algorithm for RMAB. 2) Though arms are independent in RMAB, the problem of RMAB is weakly coupled by budget constraints introduces additional difficulty.
		\item A well studied popular approach for restless bandit problem is Whittle index policy scheme, \cite{Whittle88,Gittins11}. The idea there is to map the state of each arm to a real valued number, referred to as index. Later, arms with highest indices which satisfies budget constraint are activated with different activity levels at each time slot. For $m=2,$ two action bandits, this boils down to playing arms with highest indices. 
		\item To use Whittle index policy, we require to prove indexability. This indexability approach to RMAB is non-trivial; one requires to decouple the problem into $N$ independent arms which needs relaxation of problem and Lagrangian relaxation approach. Using properties of the problem, indexability for each arm can be claimed and an index formula can derived. But to use this indexability scheme, we need more structural assumptions on transition probabilities and reward dynamics. However, an  index formula is rarely available. Often for large state space models, numerical computation of index can be  expensive. 
	\end{itemize} 
\end{remark}

The preceding discussion motivates us to look for other alternative scheme. This is studied in next section. 

\subsection{Heuristic policies} 
Here, we describe a few heuristic policies for the RMAB problem. 

\subsubsection{Myopic policy:} 
We first examine the myopic policy which is simplest heuristic policy and it has minimal computation and memory requirement. In this, the decision maker picks arms with different activity levels based on current state arms and corresponding immediate reward for different actions such that the budget constraints are satisfied. If $\mathbf{x}_i(t)$ is the state of arm $i$ at beginning of slot $t,$ the immediate reward from that state is $r_i(\mathbf{x}_i(t),a_i(t)).$ 
Arms with different activity levels are chosen according to following rule. 
\begin{eqnarray*}
\mathbf{a}_{MP}^{*}(t) \in \arg \max_{\mathbf{a} \in \overline{A}} \sum_{i=1}^{N} r_i(\mathbf{x}_i(t),a_i(t))
\end{eqnarray*}
and the highest immediate reward accrued by DM is 
\begin{eqnarray*}
	R^*(t) = \max_{\mathbf{a} \in \overline{A}} \sum_{i=1}^{N} r_i(\mathbf{x}_i(t),a_i(t)).
\end{eqnarray*}
Note that this policy does not take account of future evolution of states and  rewards, and hence this can be suboptimal.

For $m=2$ two action bandit model, in the myopic policy, $K$ arms are played at each time slot based on immediate rewards, i.e., $K$ arms with highest immediate rewards. 

\subsubsection{One step look ahead policy}
This policy is extension of myopic policy, where DM considers both immediate reward for arms and one step evolution of state and reward for selection of arms with activity level. Thus, arms are selected with different activity levels in slot $t$ according following rule. 
{\small{
\begin{eqnarray*}
\mathbf{a}_{LAP}^{*}(t) \in 	\arg \max_{\mathbf{a} \in \overline{A}} \left[ \sum_{i=1}^{N} r_i(\mathbf{x}_i(t), a_i(t)) +  
\beta \sum_{Y=(\mathbf{y}_1,\cdots, \mathbf{y}_N) \in \mathcal{S}} 
\right. \\ \left.
\prod_{i=1}^{N} P_i(\mathbf{y}_i~|~\mathbf{x}_i(t),\pi(X(t))) \max_{\mathbf{a}^{\prime} \in \overline{A}} \sum_{i=1}^{N} r_i(\mathbf{y}_i,a_i^{\prime})   \right].
\end{eqnarray*}}} 
%In this policy, $K$ arms are chosen  at each time slot based on immediate rewards plus future one step evolution of state and reward. Mathematically, this is given as follows. 
%\begin{eqnarray*}
%	\arg \max_{a \in \overline{A}} \left[ \sum_{i=1}^{N} r_i(x_i, a_i) + \beta \sum_{y=(y_1,\cdots, y_N) \in \mathcal{S}} \prod_{i=1}^{N} P_i(y_i~|~x_i,\pi(x)) \max_{a^{\prime} \in \overline{A}} \sum_{i=1}^{N} r_i(y_i,a_i^{\prime})   \right].
%\end{eqnarray*}
Observe that even one step lookahead policy is computationally challenging for large state space model.  This is due to second summation over state space. 

\subsection{Whittle index policy and  Monte Carlo approach for two action bandits}
\label{subsec:Whittleindex-singled-dim}
We now consider a special case of multi-action RMAB, i.e., two armed RMAB. Recall that action $a_i(t) = 1$ correspond to play of arm $i$ and action $a_i(t) = 0$ correspond to not play of arm $i.$ We study Whittle index policy approach. 
In this, budget constraint problem for actions, adapted for new problem with  discounted budget constraint and next using Lagrangian relaxation approach, this relaxed problem decomposed into $N$ singled armed bandit problem. This idea of constraint relaxation and Lagrangian method simplifies the problem from  $N$--armed restless bandit to $N$ single armed restless bandits.  While doing this, subsidy $W$ introduced for single-armed restless bandit problem (SARB). The dynamic program for arm $i$ with subsidy $W$ is  given by 
\begin{eqnarray*}
V_i(\mathbf{x}_i) = \max_{a_i \in \{0,1\} }
 \left\{ r_i(\mathbf{x}_i,a_i) + W(1-a_i) + 
 \right. \\ \left.
 \beta \sum_{\mathbf{y} \in \mathcal{S}_i}\left[  
  P_i(\mathbf{y}_i~|~\mathbf{x}_i,a_i) V_i(\mathbf{y}_i) \right] \right\}
\end{eqnarray*}

As discussed in the preceding section, each arm problem boils to solving simple MDP problem, but  difficulty is due to large state space. We first discuss here few special cases with single dimensional state space and later consider multi-dimensional state space. 

\subsubsection{Single dimensional state space model}
We make following assumption. 
\begin{assumption}
\begin{enumerate}
	\item The  state space of arm $i$ is single dimensional, it is denoted by $x_i$ and $x_i \in \mathcal{S}_i \subset \mathbb{Z}.$ This is true for all arms. 
	\item  The optimal policy for a single-armed bandit, i.e., for arm $i$ is threshold type. 
\end{enumerate}		
\end{assumption}
\begin{definition}[Threshold type policy]
For two action single armed bandit $i$ with a single-dimensional state space $\mathcal{S}_i$, the optimal policy is of a threshold type if 
%one of the following condition holds. 
%\begin{itemize}
%	\item Optimal action $a^*(x_i) = 1$ for all $x_i \in S_i.$
%	\item Optimal action $a^*(x_i) = 0$ for all $x_i \in S_i.$
%	\item
	 There exists $\widetilde{x}_i \in \mathcal{S}_i$ such that optimal action of arm $i,$ for given state $x_i \in \mathcal{S}_i$ satisfy 
	\begin{eqnarray*}
	a_i^{*}(x_i) = \begin{cases}
		1 & \mbox{If $x_i> \widetilde{x}_i,$} \\
		0 & \mbox{If $x_i \leq \widetilde{x}_i.$} 
	\end{cases}	
	\end{eqnarray*}
	  
%\end{itemize}
\end{definition}

%We assume that We next assume that there exists a threshold type policy and suppose $\widetilde{x}_{i}$ is a threshold for arm $i.$

The threshold type policy for arm $i$  implies that 
\begin{eqnarray*}
V_i(x_i) =
\begin{cases} 
 r_i(x_i,a_i =1) +   \\ 
  \beta  \sum_{y \in \mathcal{S}_i}  P_i(y_i~|~x_i,a_i=1) V_i(y_i) & \mbox{If $x_i > \widetilde{x}_{i},$} \\ 
 r_i(x_i,a_i=0) + W +   \\
\beta 
\sum_{y \in \mathcal{S}_i} P_i(y_i~|~x_i,a_i = 0) V_i(y_i)  & \mbox{If $x_i \leq \widetilde{x}_{i}.$}
\end{cases}
\end{eqnarray*}
Define
\begin{eqnarray*}
\widetilde{V}_{i}(x_i,a_i= 1) &:=& r_i(x_i,a_i =1) + \beta  \\ & &
\sum_{y \in \mathcal{S}_i}  P_i(y_i~|~x_i,a_i=1) V_i(y_i), \\
\widetilde{V}_{i}(x_i,a_i= 0) &:=& r_i(x_i,a_i=0) + W +  \beta  \\ & &
\sum_{y \in \mathcal{S}_i} P_i(y_i~|~x_i,a_i = 0) V_i(y_i)
\end{eqnarray*}
Then the dynamic program is given by 
\begin{eqnarray*}
V(x_i) &:=& \max\{ \widetilde{V}_{i}(x_i,a_i= 1), \widetilde{V}_{i}(x_i,a_i= 0) \}
\end{eqnarray*}

\textbf{Significance of threshold type policy:} 
A threshold type policy indicates that state space $\mathcal{S}_i$ for arm $i$ is divided into two region. Let $U_{i,0}$ and $ U_{i,1}$  be two regions and it is as follows. 
\begin{eqnarray*}
U_{i, 1} &:=& \left\{ x_i \in \mathcal{S}_i: 
\widetilde{V}_{i}(x_i,a_i= 1) > \widetilde{V}_{i}(x_i,a_i= 0)
\right\}, \\
U_{i, 0} &:=& \left\{ x_i \in \mathcal{S}_i: 
\widetilde{V}_{i}(x_i,a_i= 1) \leq \widetilde{V}_{i}(x_i,a_i= 0)
\right\}.
\end{eqnarray*}
Here, $U_{i,1}$ is set of states at which optimal action is to play arm, and $U_{i,0}$  is set of states at which optimal action is not to play arm $i.$ Note that these sets  are dependent on subsidy $W$, because value functions and actions are dependent on subsidy $W.$ Thus,  
\begin{eqnarray*}
U_{i,1}(W)&=& \left\{ x_i \in \mathcal{S}_i: a_i^*(x_i,W) = 1\right\}, \\
U_{i,0}(W)&=& \left\{ x_i \in \mathcal{S}_i: a_i^*(x_i,W) \leq 1  \right\}.
\end{eqnarray*}

\begin{definition}[Indexability \cite{Whittle88}]
Indexability for each arm is defined as set of states at which not playing is optimal choice behaves  monotonically with subsidy $W$, that is, As subsidy $W$ increases from $-\infty$ to $+\infty,$  $U_{i,0}(W)$ increases from $\emptyset$ to full set $S_i.$ 
\label{def:indexability-single-dim}	
\end{definition}
This indicates that $W$ increases, the set $U_{i,0}(W)$ monotonically nondecreasing, thus $W_2 > W_1,$ then $U_{i,0}(W_1) \subseteq U_{i,0}(W_2).$
Note that if the optimal policy is of a threshold policy, then we can claim indexability for arm. Threshold policy behavior gives sufficient condition for indexability.  

% Further this set of state space with not playing optimal increases from empty set to full state space as  subsidy increase from $-\infty$ to $\infty.$ For more detail see \cite{Whittle88,Gittins11}. 

In general the proof of indexability is hard. But indexability can be proved with additional structural assumptions on the model, transition probabilities and reward dynamics.
%, the indexability result can be proved. 

\begin{definition}[Whittle index \cite{Whittle88}]
The Whittle index of arm $i$ is the minimum subsidy $W$ at which both actions playing of arm and not playing of arm are equally good for the given state. That is,  
\begin{eqnarray*}
W(\widetilde{x_i}) = \inf \left\{W \in \mathbb{R}:  \widetilde{x_i} \in U_{0,1}(W) \right\}.
\end{eqnarray*}
\end{definition} 

Thus the index $W(\widetilde{x}_{i})$ for arm $i$ at $\widetilde{x}_{i}$ is obtained by solving following equation for $W.$  
\begin{eqnarray*}
\widetilde{V}_{i}(x_i,a_i= 0,W) - \widetilde{V}_{i}(x_i,a_i= 1,W)= 0
\end{eqnarray*} 
Then,
\begin{eqnarray*}
W +   r_i(\widetilde{x}_{i},a_i=0) + \beta 
\sum_{y \in \mathcal{S}_i} P_i(y_i~|~\widetilde{x}_{i},a_i = 0) V_i(y_i,W) \\ 
- 	r_i(\widetilde{x}_{i},a_i =1) - \beta  \sum_{y \in \mathcal{S}_i}  P_i(y_i~|~\widetilde{x}_{i},a_i=1) V_i(y_i,W)  = 0. 
\end{eqnarray*} 
%In other word, this is nothing but $\widetilde{V}_{i}(\widetilde{x}_i,a_i= 1,W)  -\widetilde{V}_{i}(\widetilde{x}_i,a_i= 1,W) =0.$  
Solving  this expression is non-trivial in many cases, because  $V_i(y_i,W)$  depends on subsidy $W$ and for large state space this recursive computation is hard. 

%To show dependence of $V_i(y_i)$ on subsidy $W,$ we rewrite the preceding expression as follows. 
%\begin{eqnarray*} 
%	W(\widetilde{x}_{i}) = r_i(\widetilde{x}_{i},a_i =1) + \beta  \sum_{y \in \mathcal{S}_i}  P_i(y_i~|~\widetilde{x}_{i},a_i=1) V_i(y_i,W) \\ 
%	-  r_i(\widetilde{x}_{i},a_i=0) - \beta 
%	\sum_{y \in \mathcal{S}_i} P_i(y_i~|~\widetilde{x}_{i},a_i = 0) V_i(y_i,W) 
%\end{eqnarray*} 
\textbf{Index computation using Monte Carlo rollout policy:} 
Note that computing $ \sum_{y \in \mathcal{S}_i}  P_i(y_i~|~\widetilde{x}_{i},a_i=1) V_i(y_i,W)$ is difficult but  we can utilize Monte Carlo rollout policy and parallel rollout policy approach developed for MDP in Section~\ref{sec:rollout-MDP}. This effectively provides way to finds approximate index. This can also increase speed of computation of index. The detail is given in Algorithm~\ref{algo:Whittle-index-compute}. The convergence of algorithm is justified using two-timescale stochastic approximations. In this algorithm, the $W$ is updated on slow timescale and Monte-Carlo rollout policy performed on fast timescale. $W$ is updated iteratively slow time scale, that means, fixed $W$ and then perform Monte Carlo rollout policy and compute approximate value function for both actions starting from given state. If difference between approximate value function under two action is less than $\epsilon,$ the index is obtained. Otherwise, change $W$ to new value and perform the Monte Carlo rollout policy.  Thus the Monte carlo policy is updating at faster timescale. The convergence of two timescale algorithm is proved in \cite{Borkar08} Chapter $6.$

%In \cite[Chapter $6$]{Borkar08}, two timescale stochastic approximation is discussed in great detail. 

\begin{algorithm}
	\caption{Whittle index computation algorithm for arm $i$}
	\begin{algorithmic}
		\STATE \textbf{Input: State of arm} $i:$ $\widetilde{x}_i$ \\
		\STATE \textbf{Initialize} $W_{old}$ 
		\STATE \textbf{1. Define: } $W_{new} = W_{old}$
		\STATE \textbf{2. Use Monte Carlo rollout policy} 
		\STATE \hspace{1cm} \textbf{Compute:} $\widetilde{V}_{i}(x_i,a_i= 1,W_{new})$ and $\widetilde{V}_{i}(x_i,a_i= 0,W_{new})$  
		\STATE \textbf{3. Define} $\Delta(x_i,W_{new}) =\widetilde{V}_{i}(x_i,a_i= 1,W_{new}) - \widetilde{V}_{i}(x_i,a_i= 0,W_{new}) $
		\STATE \textbf{4. If} $\Delta < \epsilon$ \textbf{then} 
		  \STATE \hspace{1cm} \textbf{Index: } $W(\widetilde{x}_i) = W_{new}$ and Exit 
		 \STATE \hspace{0.4cm} \textbf{Else}
		 \STATE \hspace{1cm} $W_{old} = W_{new}$
		 \STATE \hspace{1cm} $W_{new} = W_{old} + \gamma \Delta(x_i,W_{new})$
		 \STATE \hspace{1cm} \textbf{Go to step $1$}
		 \STATE \hspace{0.4cm} \textbf{End}
		 \STATE \textbf{5. Output:} $W(\widetilde{x}_i)$ 
	\end{algorithmic}
	\label{algo:Whittle-index-compute}
\end{algorithm}

\textbf{Monte Carlo rollout policy for structured MDPs: } 
We assumed threshold policy for each restless bandit. This implies that Monte Carlo rollout policy is implemented for structured MDP. For computation of index, we further treat the current state as threshold and evaluate rollout policy in which base policy $\pi$ is used. The policy $\pi$ is threshold policy with known threshold $\widetilde{x}_i.$ Thus we consider $\pi(x_i) =1$ for $x_i > \widetilde{x}_i$ and $\pi(x_i) =0$ for $x_i \leq  \widetilde{x}_i.$  In more detail, this is given in Algorithm~\ref{algo:Monte-Carlo Rollout Policy}. 

\begin{algorithm}
	\caption{Monte Carlo rollout policy for arm $i$}
	\begin{algorithmic}
		\STATE \textbf{Input: State of arm $i:$ $\widetilde{x}_i$ and $W$ }\\
		\STATE \textbf{1. Threshold policy} $\pi$ with  known  threshold $\widetilde{x}_i$
		\STATE \hspace{1cm} \textbf{If} $x>\widetilde{x_i}$ \textbf{Then} 
		\STATE  \hspace{2cm} $\pi(x) =1$  for $x \in \mathcal{S}$
		\STATE \hspace{1cm}  \textbf{Else} ($x\leq\widetilde{x_i}$) 
		\STATE  \hspace{2cm} 
		$\pi(x) =0$  for $x \in \mathcal{S}$
		\STATE \hspace{1cm} \textbf{End}
		\STATE \textbf{2. Monte Carlo rollout policy} 
		\STATE \hspace{1cm} \textbf{Compute:} For $l =1, 2, \cdots, L$
		\begin{eqnarray*}
		Q_{l,\tau}(\widetilde{x}_i, W,\pi) = \sum_{t=0}^{\tau-1} \beta^{t} \widetilde{r}(x_{i,t}, W, \pi(x_{i,t}))
		\end{eqnarray*}
		\STATE \hspace{3.5cm}   $\widetilde{r}(x_{i,t}, W, \pi(x_{i,t}) =1) := r(x_{i,t},  1) $ 	
		\STATE \hspace{3.5cm}   $\widetilde{r}(x_{i,t}, W, \pi(x_{i,t}) =0) := r(x_{i,t},  0) + W $ 	 
		\STATE \hspace{1cm} \textbf{Estimate}
		\begin{eqnarray*}
		\widetilde{Q}_{L,\tau}(\widetilde{x}_i, W,\pi) 	 = \frac{1}{L} \sum_{l=1}^{L} Q_{l,\tau}(\widetilde{x}_i, W,\pi) 
		\end{eqnarray*} 
		\STATE \hspace{1cm} \textbf{Obtain}
		\STATE \hspace{2cm} $\widetilde{V}_{i}(x_i,a_i= 1,W) = r(x_i,a_i=1) + \widetilde{Q}_{L,\tau}(\widetilde{x}_i, W,\pi)$
		
		\STATE \hspace{2cm} $\widetilde{V}_{i}(x_i,a_i= 0,W) =  r(x_i,a_i=0) + W +  \widetilde{Q}_{L,\tau}(\widetilde{x}_i, W,\pi)$ \\
		
		\STATE \textbf{3. Output:}  $\widetilde{V}_{i}(x_i,a_i= 0,W) $ and 
		$\widetilde{V}_{i}(x_i,a_i= 1,W).$			  
	\end{algorithmic}
	\label{algo:Monte-Carlo Rollout Policy}
\end{algorithm}

We next provide concentration inequality based result for rollout policy using Hoeffding inequality, see \cite{Boucheron12}. We provide bound on number of sample of trajectories needed for given approximation error  to value function under policy $\pi.$ 
\begin{theorem}
For given $\epsilon, \delta > 0$ and  $\tau$ there is $L > \widetilde{L}$ such that with probability $1-\delta$  we have 
\begin{eqnarray*}
	\bigg \vert V_{i,\pi}(\widetilde{x}_i) - \frac{1}{L} \sum_{l=1}^{L} Q_{l,\tau}(\widetilde{x}_i, \pi) \bigg \vert \leq  \epsilon.
\end{eqnarray*}
Here,
\begin{eqnarray}
L := \frac{2 \epsilon^2 (1-\beta^2)}{(2R_{\max}-R_{\min})^2(1-\beta^{\tau})\log(2/\delta)}.
\label{eqn:bound-L}
\end{eqnarray}
\end{theorem}

\begin{IEEEproof}
	Let $V_{i,\pi}(x_i)$ be the value function for arm $i$ under policy $\pi,$ and $Q_{l,\tau}(\widetilde{x}_i, \pi)$ is discounted cumulative reward along the trajectory $t=l$ with horizon length $\tau.$ This given by  
	\begin{eqnarray*}
		Q_{l,\tau}(\widetilde{x}_i, \pi) = \sum_{t=0}^{\tau-1} \beta^{t} \widetilde{r}(x_{i,t}, \pi(x_{i,t}))
	\end{eqnarray*} 
	where $x_{i,0} = \widetilde{x}_i,$ $\widetilde{r}(x_{i,t}, \pi(x_{i,t}) = 0) = r(x_{i,t}, 0) + W,$ and $\widetilde{r}(x_{i,t}, \pi(x_{i,t}) = 1) = r(x_{i,t}, 1).$  Let assume that $2R_{\min} \leq r(x,a) \leq R_{\max}.$ for all $(x,a) \in\mathcal{S}\times \mathcal{A}.$ Also assume that $R_{min}\leq W \leq R_{\max}$ Note that $\{Q_{l,\tau}(\widetilde{x}_i, \pi)\}_{l=1}^L$ are independent random trajectories generated using policy $\pi$ and $Q_{l,\tau}(\widetilde{x}_i, \pi) \in \left[\frac{R_{\min}(1-\beta^{\tau})}{1-\beta}, \frac{2R_{\max}(1-\beta^{\tau})}{1-\beta}\right].$
	Then from Hoeffding inequality (see Appendix~\ref{app:lemma-Hoeffding-inequality}), we obtain 
	
	\begin{eqnarray*}
		\prob{\bigg \vert V_{i,\pi}(\widetilde{x}_i) - \frac{1}{L} \sum_{l=1}^{L} Q_{l,\tau}(\widetilde{x}_i, \pi) \bigg \vert > \epsilon}  \leq 
		\\
		 2 \exp\left(\frac{-2 \epsilon^2(1-\beta)^2}{L(2R_{\max}-R_{\min})^2(1-\beta^{\tau})^2}\right).
	\end{eqnarray*}
We want right side in preceding equation to be small, say $\delta.$ Thus,
\begin{eqnarray*}
2 \exp\left(\frac{-2 \epsilon^2(1-\beta)^2}{L(2R_{\max}-R_{\min})^2(1-\beta^{\tau})^2}\right) = \delta.  
\end{eqnarray*}
After simplifying, we get the bound on $L$ and this is given in Eqn.~\eqref{eqn:bound-L}
This completes the proof. 
\qed 
\end{IEEEproof}
\begin{remark} 
Observe that for sufficiently large horizon length $\tau,$ the term $1-\beta^{\tau} \approx 1.$ Thus number of sampled trajectories needed is $L \approx  \frac{2 \epsilon^2 (1-\beta^2)}{(2R_{\max}-R_{\min})^2\log(2/\delta)}.$ The proposition measures the goodness of Monte Carlo rollout policy w.r.t. policy $\pi$ and provides approximation to value function.   
\end{remark}

\subsubsection{Multi-dimensional state space model} 
We now study multi-dimensional state space model for RMAB. 
\begin{assumption}
	\begin{enumerate}
		\item Assume that the state space of each arm in RMAB is multi-dimensional. That is,  $\mathbf{x}_i \in \mathcal{S} \subset \mathbb{Z}^n$ for arm $i.$ This is assumed for all arms. 
		\item The optimal policy for each single-armed restless bandit is threshold-type. 
	\end{enumerate}
\end{assumption}

Definition of threshold type policy for multi-dimensional state space model is non-trivial. It is not going to be single point anymore but it is  collection points, $\mathbf{x}_i$s.  This is called as threshold region.  We can compare $\mathbf{x}_i, \& \mathbf{y}_i \in \mathcal{S}_i$ are compared using partial order, i.e., $\mathbf{x}_i \geq \mathbf{y}_i$ iff $\mathbf{x}_i(k) \geq  \mathbf{y}_i(k)$ for $k =1,2,\cdots,n$

\begin{definition}[Threshold type policy]
	For two action single armed bandit, say arm $i$  with multi-dimensional state space $\mathcal{S}_i$, the optimal policy is of a threshold type if 
	there exists set $\Gamma_i \subset \mathcal{S}_i$ such that optimal action of arm $i,$ for given state $\mathbf{x}_i \in \mathcal{S}_i$ satisfy  
	\begin{eqnarray*}
		a_i^{*}(\mathbf{x_i}) = \begin{cases}
			1 & \mbox{if $\mathbf{x}_i> \mathbf{y}_i,$  $ \mathbf{y}_i \in \Gamma_i,$  } \\
			0 & \mbox{if $\mathbf{x}_i \leq \mathbf{y}_i,$ $ \mathbf{y}_i \in \Gamma_i.$} 
		\end{cases}	
	\end{eqnarray*}
 	
\end{definition}
Note that $\Gamma_i$ is threshold-type region, for which both actions are optimal but for our purpose we choose not playing as optimal action.  Moreover, this set  for arm $i$ is given as follows. 
\begin{eqnarray*}
	\Gamma_i = \left\{ \mathbf{x}_i \in \mathcal{S}_i :
	 r_i(\mathbf{x}_i,a_i=0) + W + 
	\right. \\ \left.  
	  \beta 
	\sum_{\mathbf{y}_i \in \mathcal{S}_i} P_i(\mathbf{y}_i~|~\mathbf{x}_i,a_i = 0) V_i(\mathbf{y}_i) \geq   r_i(\mathbf{x}_i,a_i =1) +
	\right. \\ \left. 	
	 \beta  \sum_{\mathbf{y}_i \in \mathcal{S}_i}  P_i(\mathbf{y}_i~|~\mathbf{x}_i,a_i=1) V_i(\mathbf{y}_i) 
	  \right\}
\end{eqnarray*}
This is set of integers that divides state space into two region for  $\mathcal{S} \subset \mathbb{Z}^n.$ To have the threshold type property for each arm, we need structure on transition probabilities and reward probabilities. This we will not discuss here but we assume these are satisfied. 
The more discussion on assumptions for threshold type policy and necessary conditions are given in \cite[Chapter $4$ Section $4.7.3$]{Puterman14}, and  details on integer lattices and submodularity is in \cite{Topkis78,Topkis98}.  

\textbf{Indexability of an arm:} As studied in section~\ref{subsec:Whittleindex-singled-dim}, define  $U_{i,1}$ is set of states at which optimal action is to play arm, and $U_{i,0}$  is set of states at which optimal action is not to play arm $i.$ These sets  are dependent on subsidy $W.$ Thus,  
\begin{eqnarray*}
	U_{i,1}(W)&=& \left\{ \mathbf{x}_i \in \mathcal{S}_i: a_i^*(\mathbf{x}_i,W) = 1\right\}, \\
	U_{i,0}(W)&=& \left\{ \mathbf{x}_i \in \mathcal{S}_i: a_i^*(\mathbf{x}_i,W) \leq 1  \right\}.
\end{eqnarray*}

From Definition~\ref{def:indexability-single-dim} of indexability,  one requires to show that as subsidy $W$ increases from $-\infty$ to $+\infty$  $U_{i,0}(W)$ increases from $\emptyset$ to full set $\mathcal{S}_i.$  As mention earlier, threshold policy provides sufficient condition for indexability, and this need structure on problem. 

\textbf{Whittle index computation:} As in general obtaining closed form expression for value function is difficult, we use Monte-Carlo rollout policy as studied in  section~\ref{subsec:Whittleindex-singled-dim}. We can has similar numerical algorithm, but here complexity is more due to multi-dimensional state space model.  

\subsection{Whittle index policy and  Monte Carlo approach for multi-action bandits}
We now study multi-action RMAB, i.e., number of actions $m>2.$ Suppose that the state space is single dimension, i.e.,  $\mathcal{S}_i \subset \mathbb{Z}.$ A state in slot $t$  is $x_i(t)$ for arm $i.$  Recall that action space of arm $i$ is $\mathcal{A}_i,$ action of arm $i$ in slot $t$ is $a_i(t).$  For simplicity,  we assume that all arms has same number of actions, $\vert \mathcal{A}_i \vert =  M $ and $\mathcal{A}_i = \{ 1,\cdots, M \}$ for all $i.$

Essential idea of using Whittle index appproach is to decompose the problem into single-armed restless bandit, this is done using Lagrangian relaxation method, and budget constraint of actions brought  into objective function. For single-armed restless bandit, the dynamic program with Lagrangian multiplier is given by
\begin{eqnarray}
	V_i({x}_i) = \max_{a_i \in \mathcal{A}_i }
	\left\{ r_i({x}_i,a_i) + W (M-a_i) + \nonumber 
	\right. \\ \left. 
	 \beta \sum_{{y} \in \mathcal{S}_i}
	P_i({y}_i~|~{x}_i,a_i) V_i({y}_i) \right\},
	\label{eqn:dynamic-prog-multi-action}
\end{eqnarray}
Note that the Lagrangian multiplier $W$ is subsidy provided per unit time per unit resource consumption. 

\begin{remark}
	Observe from Eqn.~\eqref{eqn:dynamic-prog-multi-action}, that
	$W(M-a_i)$ is decreasing in activity level $a_i$ 
	Hence as the number activity level increases, i.e., $a_i$ increases, the total subsidy $\widetilde{W}(a_i) := W(M-a_i)$ obtained from  the higher activity level decreases. For highest activity level, the subsidy is zero. Thus, we have $\widetilde{W}= 0$ for $a_i = M.$ Also, $\widetilde{W}_1< \widetilde{W}_2$ whenever $a_{i,2} < a_{i,1}.$  Here, $a_{i,j}$ for $j=1,2$ indicates the two different activity level for arm $i.$
\end{remark}

The next objective is to define the indexability and Whittle index for multi-action bandit model. The idea is to extend notion of indexability of two action single armed bandit model to multi-action single armed bandit model. 

\textbf{Indexability for multi-action bandit:} 
The definition of indexability will depends on current state and activity level fixed by DM for arm $i.$ 
For given subsidy $W$ and fix activity level $\alpha_i \in \mathcal{A}_i,$  define  
\begin{eqnarray*}
V_i(W,\alpha_i) := \{x_i \in \mathcal{S}_i: a^{*}(W,x_i) \leq \alpha_i \}.
\end{eqnarray*}
It is the collection of states for which optimal action is chosen less than equal to fixed activity level $\alpha_i.$ 

We  observe from dynamic program in Eqn.~\eqref{eqn:dynamic-prog-multi-action}, that if  total subsidy $\widetilde{W}$ increases, then the expected reward is high for low activity level.  
As subsidy $W$ increases, total subsidy increases. This suggests that the optimal action chpsen is low level activity. Further,  this implies that  the states for which optimal actions is less than fixed activity level is increases. In other word $V_i(W,\alpha_i)$ increases with in increase $W.$ This is notion used for definition of indexability, which is also referred to as Full indexability in \cite{Glazebrook11,Hodge15}.  This notion will be used in our work. 

\begin{definition}[Full indexability]
	The arm $i$ is called full indexable if $V_i(W,\alpha_i)$ is non decreasing in $W$ for each  $\alpha_i \in \mathcal{A}_i.$  
\end{definition}    

If all arms are full indexable, then RMAB is called full indexable. 

\begin{definition}[Whittle index for multi-action bandit]
	The Whittle index for arm $i$ is defined as follows. $W_{i}: 
	\mathcal{S}_i \times \mathcal{A}_i \rightarrow \mathbb{R}$ and \begin{eqnarray*}
	W_i(x_i,\alpha_i) = \inf\left\{ W \in \mathbb{R}:  x_i \in V_i(W,\alpha_i)   \right\} .
	\end{eqnarray*}
\end{definition}
The following result and which is highly intuitive.
\begin{lemma}
	$W_i(x_i,\alpha_i)$ is decreasing in $\alpha_i$ for fixed $x_i.$  
	\label{lemma:W-decrease-alpha}
\end{lemma}
\begin{IEEEproof}
	If arm is fully indexable, then the Whittle index is minimum amount of subsidy required such that optimal actions or activity level less than $\alpha_i$ for given state $x_i$ and activity level $\alpha_i.$ It is the subsidy at arm $i$ for raising activity level $\alpha_i$ to $\alpha_i +1$ for given state $x_i.$ The subsidy is less than $W_i(x_i,\alpha_i),$ that means reward from low activity level is less. Hence higher activity levels are preferable. If the subsidy is higher than index  $W_i(x_i,\alpha_i)$ then higher activity level are not preferable. One can define $W_{i}(x_i, \alpha_i = M) = 0$ for all $x_i \in \mathcal{S}_i.$  Then as activity level increases  for fixed $x_i$ from $\alpha_i$ to $\alpha_i+1$ using Eqn.~\eqref{eqn:dynamic-prog-multi-action}, and this discussion it is clear that $W_i(x_i,\alpha_i)$ is decreasing in activity level $\alpha_i.$
\end{IEEEproof}

\textbf{Idea on proof of full indexability:}
 We have to show full indexability. This can be often shown when the optimal policy is monotone in action space. To claim the monotone policy result, we need more structural assumption on reward dynamic and transition probabilities, see \cite[Chapter $6,$ Theorem $6.11.7$]{Puterman14}. Assuming these structural assumptions,  the optimal stationary policy is monotone. This is nothing but threshold policy with $M-1$ threshold and this divides the state space into $M$ regions, in each region of state space, only one action is optimal. This allows to claim full indexability result.

\textbf{Index computation for multi-action bandit: }Under full indexability of arm, we  obtain index $W_i$ for each arm  $1\leq i \leq N.$ This raises following questions which were not  in two-action bandit model. 
\begin{itemize}
	\item What are the activity level being chosen for each arm?
	\item How to meet budget constraint for activity levels?
\end{itemize}
We answer this by index computation with greedy heuristic algorithm. This is motivated from~\cite{Glazebrook11}. 

In the index policy, say $\pi(W)$ with fixed subsidy $W,$ selection of optimal activity level for each arm satisfies the following property.
\begin{eqnarray*}
\mathbf{a}^*(W,\mathbf{x}) = \mathbf{a}^* &\mbox{iff $W_i(a_i^*-1,x_i) > W > W_i(a_i^*,x_i)$ for all $i.$}
\end{eqnarray*}  
It follows from~Lemma~\ref{lemma:W-decrease-alpha}. Using this, one can consider the greedy heuristic algorithm.

In this algorithm first step is to start with initial activation level to zero for all arms. Clearly, this meets the budget constraint. In the second step  the arm with highest Whittle index can be figure out for given allocation (activity level). Thus 
\begin{eqnarray*}
i^* =  \arg \max_{1 \leq i \leq N} W_i(x_i, \alpha_i)
\end{eqnarray*}
If there are more than one arm with highest index, then arm will be picked randomly. 
In the third step, increase the activity level for arm with highest index earlier, this increase in activity level by $1$ unit, Later verify the budget constraint if this satisfies, $\sum_{i=1}^{N}a_i \leq K,$ then again repeat second step for this new allocation.  Using this we again pick arm with highest index and change their allocation strategy. This process is repeated until constraint satisfy $\sum_{i=1}^{N}a_i > K.$

Recall that two-action model we have nice structure which allowed us to compute the index numerically or using simulation method. 
In this case, how can we use structural result to come up with index for each arm either numerically or simulation based approach. 

We need to compute the index for each state and activity level for given arm. The computation of index  is non-trivial. As subsidy increases, the low level activity becomes more rewarding, that is optimal action is to choose low level activity. In the index, for each activity level $\alpha_i$ we have to compare the value function for action $\alpha_i$ and $\alpha_i-1$ with given state. Thus the  index computation is solution of following equation.  
\begin{eqnarray*}
r_i({x}_i,\alpha_i) + W (M-\alpha_i) + \beta \sum_{{y} \in \mathcal{S}_i}
P_i({y}_i~|~{x}_i,\alpha_i) V_i({y}_i) -  \\
\left[ 
 r_i({x}_i,\alpha_i-1) + W (M-\alpha_i + 1 ) + 
 \right. \\ \left. 
  \beta \sum_{{y} \in \mathcal{S}_i}
P_i({y}_i~|~{x}_i,\alpha_i-1) V_i({y}_i) \right] = 0
\end{eqnarray*}
In this computation, multi-timescale stochastic approximation algorithm is used where in  slower timescale, the subsidy $W_t$ iteration is run and fast timescale the value iteration algorithm is performed. This gives  
\begin{eqnarray*}
W_{t+1}  &=& W_t + \gamma_{t+1} \Delta W_{t+1}, 
\end{eqnarray*}
\begin{eqnarray*}
V_{i,t+1}({x}_i) = \max_{a_i \in \mathcal{A}_i }
\left\{ r_i({x}_i,a_i) + W_{t} (M-a_i) + 
\right. \\ \left. 
\beta \sum_{{y} \in \mathcal{S}_i}
P_i({y}_i~|~{x}_i,a_i) V_{i,t} ({y}_i) \right\},
\end{eqnarray*}
where
\begin{eqnarray*}
\Delta W_{t+1} =r_i({x}_i,\alpha_i-1) + \beta \sum_{{y} \in \mathcal{S}_i}
P_i({y}_i~|~{x}_i,\alpha_i-1) V_{i,t}({y}_i) - \\  r_i({x}_i,\alpha_i) 
 - \beta \sum_{{y} \in \mathcal{S}_i} P_i({y}_i~|~{x}_i,\alpha_i) V_{i,t}({y}_i).
\end{eqnarray*}
Convergence of multiple timescale algorithms is discussed in \cite[Chapter $6$]{Borkar08}. Idea there is to show that this algorithm and solution of singularly perturbed differential equations asymptotically converges. 

Even in this case computation of $ \sum_{{y} \in \mathcal{S}_i} P_i({y}_i~|~{x}_i,\alpha_i) V_{i,t}({y}_i)$  can be infeasible. Hence we can utilize  Monte Carlo based rollout policy for this approximation, as given in section~\ref{subsec:Whittleindex-singled-dim}.

Analogously, we can  extend this for multi-dimensional state space, but complexity of problem increases and need much more structure on the problem  to claim monotone policy and  hence  indexability.  

In next section we discuss a simple Monte Carlo rollout policy without index approach. 

\section{Monte Carlo rollout policy for non-indexable two-action bandits} 
\label{sec:RMAB-rollout-nonindexable-two-action}
Most often showing indexability is challenging task.  RMAB is shown to be indexable under special structure on model or for special cases. This motivate  to look the problem  directly by Monte Carlo rollout policy without  indexability approach.  

There are few differences between rollout policy with index based approach and directly applying rollout policy to RMAB without indexability.  
In preceding section for RMAB problems, we assumed threshold policy result and note that  to establish this policy one require sufficient structure on the problem. Later, we used threshold policy result in rollout policy and that is done separately for each arm. Finally, we computed approximate index, see Algorithm~\ref{algo:Whittle-index-compute} and~\ref{algo:Monte-Carlo Rollout Policy}. 
In RMAB without indexability approach, we do not consider this separation of arms. Instead, we directly simulate the rollout policy, where $K$ arms are played each time instant in trajectory with highest immediate reward for given current state. This is one such simple policy that we used. There can be other variant of this policy and this will be discussed later.

%This is  simulation based approch, where $L$  number of simulated trajectories starting from state $(x,a)$ are generated using some fixed policy $\pi$ and these simulations are run upto depth of horizon $T.$ In each stage, the empirical discounted reward obtained from following policy $\pi$ is backed up and the $K$ arms are choosen using this estimate of state action value function.   In this model, next state from given state and action according to state transition dynamic $P(y~|~x,a).$ This is  generative model approach, where next state transition and rewards are generated using model. Moreover this reduces computation.   The detail procedure is explained below. 

Let $Q^{l}_{\pi,t}(\mathbf{x},\mathbf{a},\mathbf{y}_{t,l},\pi(\mathbf{y}_{t,l}))$  be the state action value for $l$th sample at stage $t$ under policy $\pi$ when initial state and action is $(\mathbf{x},\mathbf{a})$ and current state is $\mathbf{y}_{t,l}$ for $l$th sample and policy is $\pi(\mathbf{y}_{t,l}).$ Since this is non-indexable approach, we do not have subsidy $W.$ We consider the policy $\pi$ that selects the arms in current state with maximum immediate reward, thus $\pi(\mathbf{y}_{t,l}) = \max_{a \in \overline{A}} \sum_{i=1}^{N} r_i(y_{i,t,l},a_i).$
In our RMAB problem with policy $\pi(x),$ $K$ arms are played at each instant, and the reward is obtained only from $K$ played arms and no reward from not played arms. Hence $Q^{l}_{\pi,t}(\mathbf{x},\mathbf{a},\mathbf{y}_{t,l},\pi(\mathbf{y}_{t,l}))$ is sum of reward from $K$ played arms in stage $t,$ sample $l$ and state $\mathbf{y}_{t,l}$ under policy $\pi,$ i.e.,  $Q^{l}_{\pi,t}(\mathbf{x},\mathbf{a},\mathbf{y}_{t,l},\pi(\mathbf{y}_{t,l})) = \max_{a \in \overline{A}} \sum_{i=1}^{N} r_i(y_{i,t,l},a_i).$

For $\tau=1,$ this  rollout MC policy is a simple approximation to one-step lookahead policy and it given as follows. 
\begin{eqnarray*}
	\widetilde{Q}_{\pi,1}(\mathbf{x},\mathbf{a}) &=& \sum_{i=1}^{N} r_i(x_i,a_i) + \\ & & \beta\left[    \frac{1}{L} \sum_{l=1}^{L}   Q_{\pi,1}^{l}(\mathbf{x},\mathbf{a},y_{1,l}, \pi(y_{1,l})) \right] 
\end{eqnarray*}
and 
\begin{eqnarray*}
	\pi^{\prime}(\mathbf{x}) \in \arg \max_{\mathbf{a} \in \overline{A}} \widetilde{Q}_{\pi,1}(\mathbf{x},\mathbf{a}).
\end{eqnarray*}

A variant of this policy is for $\tau>1,$ in that case there is set of  trajectories of state-action-and reward according to policy $\pi$ starting state and action $(x,a)$ upto length of horizon $\tau.$ There are such $L$ trajectories are generated using generative model. This is averaged over number of trajectories. Mathematically, $\{Q^{l}_{\pi,t}(x,a,y_{t,l},\pi(y_{t,l}))\}_{t=1}^{\tau}$ is a trajectory of state action value from state-action $(x,a)$ for $l$th sample. Then, total discounted reward for $l$th trajectory running upto horizon length $\tau$ with policy $\pi$ is 
\begin{eqnarray*}
	\overline{Q}_{\pi, \tau}^{l}(\mathbf{x},\mathbf{a}) 
	= \sum_{t=1}^{\tau} \beta^{t-1} Q_{\pi,t}^{l}(\mathbf{x},\mathbf{a},\mathbf{y}_{t,l}, \pi(\mathbf{y}_{t,l})).
\end{eqnarray*}
The state action value estimate from this policy is 
\begin{eqnarray*}
	\widetilde{Q}_{\pi,T}(\mathbf{x},\mathbf{a}) &=& \sum_{i=1}^{N} r_i(x_i,a_i) + \beta\left[    \frac{1}{L} \sum_{l=1}^{L}    \overline{Q}_{\pi, T}^{l}(\mathbf{x},\mathbf{a}) \right], 
\end{eqnarray*}
and policy is 
\begin{eqnarray*}
	\pi^{\prime}(\mathbf{x}) \in \arg \max_{\mathbf{a} \in \overline{A}} \widetilde{Q}_{\pi,T}(\mathbf{x},\mathbf{a}).
\end{eqnarray*}

This is a simple  rollout policy with greedy  behavior of $\pi.$ This does not take account of future reward  while playing arms.  

Another variant of policy $\pi$ is to consider stochastic policy, where we use exploration-exploitation behavior of $\pi.$ In simulated policy $\pi,$ at each instant the $K$ arms having the highest immediate rewards  are played  based on current state with probability $1-\epsilon$ and this is exploitation scheme. The exploration is performed with probability $\epsilon$ where any $K$ arms are played randomly that does not depend on state. Moreover, this exploration parameter $\epsilon$ is decreasing with time horizon $\tau$ in each trajectory.

There are other various version of stochastic policies are possible, where different weight could be assigned to different arms. While doing this knowledge of subset of state transition probabilities and rewards can be utilized.

This rollout based policy approach is feasible alternative with very least structure on dynamics of arms. 

\section{Monte Carlo rollout policy for non-indexable multi-action restless bandits}
\label{sec:RMAB-rollout-nonindexable-multi-action}
In this section we develop Monte Carlo rollout policy for non-inddexable multi-action restless bandits. As mention in preceding section, different actions for a bandit represent the activity of the arm. We now do not impose any structure on the problem and hence problem does not have threshold structure. Therefore the multi-action restless bandit may be non-indexable. 

We use Monte-Carlo rollout policy. Let $\{Q^{l}_{\pi,t}(\mathbf{x},\mathbf{a},\mathbf{y}_{t,l},\pi(\mathbf{y}_{t,l}))\}_{t=1}^{T}$ is a trajectory of state-action value from $(\mathbf{x},\mathbf{a})$ for $l$th sample. Then, total discounted reward for $l$th sample and trajectory running upto horizon length $\tau$ with policy $\pi$ is 
\begin{eqnarray*}
	\overline{Q}_{\pi, \tau}^{l}(\mathbf{x},\mathbf{a}) 
	= \sum_{t=1}^{\tau} \beta^{t-1} Q_{\pi,t}^{l}(\mathbf{x},\mathbf{a},\mathbf{y}_{t,l}, \pi(\mathbf{y}_{t,l})).
\end{eqnarray*}

The policy $\pi$ is implemented in which the arms with different activity level is selected at each instant along trajectory and this is done using greedy approach. Then 
\begin{eqnarray*}
Q_{\pi,t}^{l}(\mathbf{x},\mathbf{a},\mathbf{y}_{t,l}, \pi(\mathbf{y}_{t,l})) = 
\max_{ \pi(\mathbf{y}_{t,l}) \in \overline{A} } 
\sum_{i=1}^{N} r_i(\mathbf{y}_{i,t,l},a_i)
\end{eqnarray*} 

Here, $\overline{A} = \{\mathbf{a} = (a_1, \cdots, a_N) \in \mathcal{A} : \sum_{i=1}^{N}a_i \leq K, a_i \in \mathcal{A}_i, 1 \leq i \leq N \}.$

Using this trajectories are generated and  and we compute $\overline{Q}_{\pi, \tau}^{l}(\mathbf{x},\mathbf{a})$ for all $l.$

The state action value estimate from this policy is 
\begin{eqnarray*}
	\widetilde{Q}_{\pi,T}(\mathbf{x},\mathbf{a}) &=& \sum_{i=1}^{N} r_i(\mathbf{x}_i,a_i) + \beta\left[    \frac{1}{L} \sum_{l=1}^{L}    \overline{Q}_{\pi, T}^{l}(\mathbf{x},\mathbf{a}) \right], 
\end{eqnarray*}
and policy is 
\begin{eqnarray*}
	\pi^{\prime}(\mathbf{x}) \in \arg \max_{a \in \overline{A}} \widetilde{Q}_{\pi,T}(\mathbf{x},\mathbf{a}).
\end{eqnarray*}
%Here, $\overline{A} = $
%In this the policy $\pi$ is used for the given state 

Note that the problem with multi-action is more complex than two armed bandit for simple rollout policy with greedy algorithm for $\pi.$  This is because implementation of greedy policy with satisfying budget constraint and activity levels is non-trivial.

%In general when more than two actions are available for each restless bandit, the problem becomes more challenging. 

%\rahul{This extension is non trivial, we need more thinking and work}

%\rahul{Need to cite paper by Nino-Mora, \cite{Nino-Mora01}.}

\section{Concluding remarks and possible extensions}
\label{sec:conclusion-remark}
In this paper our objective was a develop a computational perspective for MDP, RMAB and multi-action RMAB. There is a trade off between computational time and accuracy of the solution.  

We studied a simple Monte Carlo rollout policy. This is often feasible for multi-dimensional state space in MDP when there is limited computational budget. This scheme is very useful for RMAB, particularly multi-action bandits when indexability is very difficult to claim. 

In our future work we plan to provide extensive computational results for different applications. This work gives various directions future research. We mention here few  possible extensions.  
\begin{enumerate} 
\item \textbf{Structured MDP:} We studied Monte-Carlo rollout policy for single threshold MDP. The threshold would be a curve for multidimensional MDP.  We can utilize the rollout policy algorithm for such problems. Finding a base policy can be   easy since state space, transition probabilities and rewards are ordered, i.e., partially ordered. 
\item \textbf{Importance-sampling based rollout policy:} 
The Monte Carlo rollout policy is a simulation based approach, where next state is sampled by a generator using known model with uniform sampling. In that case variance can high and to reduce the variance, Importance sampling is used. The current rollout policy can be modified for Importance sampling. 
\item \textbf{Constrained MDPs and RMAB:} 
Constrained MDPs (CMDP) are well studied \cite{Altman99}. In these problems, there are multiple objective reward functions. These follow some constrained inequality. Here, a stationary deterministic policy may not be feasible. Thus  stationary stochastic policies are well studied. Rollout policy approach for CMDP is another direction of  work. Similarly, in RMAB problems, there can be multiple objective reward functions. The extension of stochastic policy with Rollout policy approach is an open issue. 

\item \textbf{ Hidden Markov RMAB with multi-state space model:}  This is a class of partially observable RMAB, where states are not known but an observation is available from each state. Here, a belief about the state is maintained. This belief is vector and it is point in $M-1$ dimensional simplex. Here computing value function and even claiming indexability is very hard for more than two state model and two action bandits. The hidden Markov RMAB is studied for two state and two action bandits in \cite{Meshram18,Kaza19}, for multi-state in \cite{Ouyang14}  .
Rollout policy can be useful a approach with limited computation.   
%\rahul{Add few references on multi-state model}
\end{enumerate}

%\rahul{Develop some strong applications which are relevant in today time---IoT, Cyber physical systems, Security in cyber physical system, recommendation systems.} 

%\subsection{Partition of state space into $L$ region and asynchronous algorithm}
%We partition the state space into $L$ nonempty disjoint subset, let $\mathcal{S}_1,\cdots, \mathcal{S}_L$ be the partition of state space.  Another algorithm is asynchronous version of Monte-carlo rollout policy. 

\bibliographystyle{IEEEbib}
\bibliography{rmab-ref}

\appendix

\subsection{Proof of Lemma~\ref{lemma:subopt-rollout1}}
\label{app:lemma-subopt-rollout1}

The proof of this is straightforward, idea there is to run value iteration with base policy $\pi$ for finite horizon and we can get the following inequality,
\begin{eqnarray*}
	V^*(\mathbf{x}) - V_{\pi}(\mathbf{x}) \leq   V^*(\mathbf{x}) - V_{\pi,\tau}(\mathbf{x}). 
\end{eqnarray*}
preceding inequality   because  
\begin{eqnarray*}
	V_{\pi,\tau}(\mathbf{x})  &\leq&  \sum_{t=1}^{\tau} \beta^t \mathrm{E}  \left[ r(\mathbf{x}(t), \pi_{rh}(\mathbf{x})) \right] + \beta^{\tau+1} 
	\mathrm{E} \left[V_{\pi,\tau}(\mathbf{x})\right], \\
	V_{\pi,\tau}(\mathbf{x})  &\leq& V_{\pi}(\mathbf{x}).  
\end{eqnarray*}
Note that we can have following inequality, 

\begin{eqnarray*}
	\vert V^*(\mathbf{x}) - V_{\pi,\tau}(\mathbf{x}) \vert &\leq& \beta^\tau \max_{\mathbf{x}} \vert V^*(\mathbf{x})- V_0(\mathbf{x}) \vert, \\ 
	&\leq& \beta^\tau \frac{R_{\max}}{1-\beta}.  
\end{eqnarray*}
\qed 

This result is mentioned  in \cite{Lerma90}, see Theorem $3.1.$

\subsection{Proof of Lemma~\ref{lemma:error-approx-policy-pi}}
\label{app:lemma-error-approx-policy-pi}

 We obtain upper bound.  Consider 
\begin{eqnarray}
	V^*(\mathbf{x}) - V_{\pi}(\mathbf{x})  = (V^*(\mathbf{x}) - T(V)(\mathbf{x})) + 
	\nonumber \\
	 (T(V)(\mathbf{x}) -  V_{\pi}(\mathbf{x})).
	\label{eqn:value-func-diff-pi}
\end{eqnarray}
Here $T$ is Bellman operation.
We give bound on preceding expression. First term in RHS of Eqn~\eqref{eqn:value-func-diff-pi} is written as follows.   
\begin{eqnarray}
\vert V^*(\mathbf{x}) - T(V)(\mathbf{x}) \vert  \leq  \vert V^*(\mathbf{x}) - V_{H}(\mathbf{x}) \vert 
+ \nonumber \\
 \vert  V_{H}(\mathbf{x}) - T(V)(\mathbf{x}) \vert  
\label{eqn:term-diff-bound1}
\end{eqnarray}
Here $V_{H}(\mathbf{x})$ is the value function with finite horizon length $H$ and initial state $\mathbf{x}.$ 
Idea is to give bound on $\vert  V_{H}(\mathbf{x}) - T(V)(\mathbf{x}) \vert.$  This can be derived using value iteration algorithm. Suppose the  value iteration $V_t,$ $V_t := T(V_{t-1}).$ Note that operator $T$ is contraction mapping.  Thus,
\begin{eqnarray}
	\vert T(V_{H-1})(\mathbf{x}) - T(V)(\mathbf{x}) \vert &\leq& \beta \sup_{\mathbf{x} \in \mathcal{S}} \vert V_{H-1}(\mathbf{x})- V(\mathbf{x})  \vert \nonumber \\ 
	& \leq & \beta \epsilon.   
	\label{eqn:bound-term1}
\end{eqnarray}
In last inequality used assumption made in Lemma. 

Next we want bound on $	\sup_{\mathbf{x} \in \mathcal{S}} \vert V^{*}(\mathbf{x}) - V_{H}(\mathbf{x}) \vert$ in Eqn.~\eqref{eqn:term-diff-bound1} 
This is a difference between the optimal value function and finite horizon value function.Then 
\begin{eqnarray*}
	\sup_{\mathbf{x} \in \mathcal{S}} \vert V^{*}(\mathbf{x}) - V_{H}(\mathbf{x}) \vert &=& \sup_{\mathbf{x} \in \mathcal{S}} \vert T(V^{*})(\mathbf{x}) - T(V_{H-1})(\mathbf{x}) \vert \\
	&=& \beta \sup_{\mathbf{x} \in \mathcal{S}} \vert V^{*}(\mathbf{x}) - V_{H-1 }(\mathbf{x}) \vert \\
	&=& \beta^2 \sup_{\mathbf{x} \in \mathcal{S}} \vert V^{*}(\mathbf{x}) - V_{H-2 }(\mathbf{x}) \vert.
\end{eqnarray*}
Repeating this, we have 
\begin{eqnarray}
	\sup_{\mathbf{x} \in \mathcal{S}} \vert V^{*}(\mathbf{x})  - V_{H}(\mathbf{x}) \vert &=&  \beta^H \sup_{\mathbf{x} \in \mathcal{S}} \vert V^{*}(\mathbf{x}) - v_{0}(\mathbf{x}) \vert \nonumber  \\
	&\leq & \beta^H \frac{R_{\max}	}{1-\beta}.
	\label{eqn:bound-term2}
\end{eqnarray} 
From Eqn.~\eqref{eqn:bound-term1} and \eqref{eqn:bound-term2}, we  have following bound on Eqn.~\eqref{eqn:term-diff-bound1}. 
\begin{eqnarray*}
	\vert V^*(\mathbf{x}) - T(V)(\mathbf{x}) \vert \leq  \beta^H \frac{R_{\max}	}{1-\beta} + \beta \epsilon. 
\end{eqnarray*}
We next want bound on $ T(V)(\mathbf{x}) -  V_{\pi}(\mathbf{x}),$ and we can  obtain  
\begin{eqnarray*}
	T(V)(\mathbf{x}) -  V_{\pi}(\mathbf{x}) \leq \frac{\beta \epsilon (1+\beta)}{1-\beta}
\end{eqnarray*}  
for all $\mathbf{x} \in \mathcal{S}.$  Then combining all preceding inequalities, we can have 
\begin{eqnarray*}
	V^*(\mathbf{x}) - V_{\pi}(\mathbf{x}) & \leq & \beta^H \frac{R_{\max}	}{1-\beta} + \beta \epsilon + \frac{\beta \epsilon (1+\beta)}{1-\beta} \\
	&=& \beta^H \frac{R_{\max}	}{1-\beta} + \frac{2 \beta \epsilon }{1-\beta}.
\end{eqnarray*}
This gives desired result. 
We now want to show $	T(V)(\mathbf{x}) -  V_{\pi}(\mathbf{x}) \leq \frac{\beta \epsilon (1+\beta)}{1-\beta}.$ To claim this bound, there is interesting trick which is used. 
First look at $T(V)(\mathbf{x})$ and one can derive the following bound (need few steps). 
\begin{eqnarray*}
	T(V)(\mathbf{x}) \leq   \mathrm{E}\left[\sum_{t=0}^{n} \beta^t R_t(\mathbf{x}(t), \pi(\mathbf{x}(t)) ) ~\vert~\mathbf{x}_0 = \mathbf{x}\right] + \\
	 \beta^{n+1} \mathrm{E}\left[ T(V(\mathbf{x}(n+1))) ~\vert~ \mathbf{x}_0=\mathbf{x}\right] + \\
	\left[\beta \epsilon (1+\beta) + \beta^2 \epsilon (1+\beta ) + \cdots + \beta^{n+1} \epsilon 
	(1+\beta)\right].  
\end{eqnarray*}
Next letting $n\rightarrow \infty,$ first term becomes $V_{\pi}(\mathbf{x}),$ second term goes to zero and all the terms in third square bracket becomes $\frac{\epsilon (1+\beta)}{1-\beta}.$  This gives
\begin{eqnarray*}
	T(V)(\mathbf{x}) -  V_{\pi}(\mathbf{x}) \leq \frac{\beta \epsilon (1+\beta)}{1-\beta}. 
\end{eqnarray*}
This completes the proof. 
\qed 
\subsection{Proof of Lemma~\ref{lemma:large-horizon-rollout-pi}}
\label{app:lemma-large-horizon-rollout-pi}
%Note that to derive the proof of this result one requires following property. 
%\begin{property}
%	Suppose there exists $\phi \in B(\mathcal{S})$ for which $T_{\pi}(\phi)(\mathbf{x}) \geq \phi (\mathbf{x})$ for all $\mathbf{x} \in  \mathcal{S},$ then $V_{\pi}(\mathbf{x}) \geq \phi(\mathbf{x}).$ Here, $T_{\pi}$ is the monotonicity  operator under base policy $\pi.$ 	
%\end{property}

From rollout policy for MDP in Section~\ref{sec:rollout-MDP}, we can have following inequality. 
\begin{eqnarray*}
V_{\pi_{ro}, \tau}(\mathbf{x}) \geq V_{\pi, \tau}(\mathbf{x}).
\end{eqnarray*}
For given policy $\pi$ 
\begin{eqnarray*}
 V_{\pi}(\mathbf{x}) &=& V_{\pi, \tau}(\mathbf{x})  + \beta^{\tau} \expect{V_{\pi}(\mathbf{x}_{\tau})~|~\mathbf{x}} \\
 &\leq & V_{\pi_{ro}, \tau}(\mathbf{x}) + \beta^{\tau} \expect{V_{\pi}(\mathbf{x}_{\tau})~|~\mathbf{x}} \\
 &\leq & V_{\pi_{ro}, \tau}(\mathbf{x}) + \beta^{\tau} \frac{R_{\max}}{1-\beta}
\end{eqnarray*}

Thus, for letting $\epsilon >  \beta^{\tau} \frac{R_{\max}}{1-\beta}$ we get
\begin{eqnarray*}
V_{\pi_{ro}, \tau}(\mathbf{x})  &\leq& V_{\pi}(\mathbf{x}) - \epsilon.
\end{eqnarray*}

\subsection{Hoeffding inequality from \cite{Boucheron12}}
\label{app:lemma-Hoeffding-inequality}
Let $x_1, x_2, \cdots, x_t$ be independent  bounded variables such that $a_i \leq x_i \leq b_i.$ Let $S_t = \sum_{i=1}^{t} x_i.$ and $\expect{S_t}$ Then for any $\epsilon >0$ we have 
\begin{eqnarray*}
\prob{\vert S_t - \expect{S_t} \vert > \epsilon  } \leq 2 \exp\left(\frac{-2\epsilon^2}{\sum_{i=1}^{t}(b_i-a_i)^2}\right)
\end{eqnarray*} 

\subsection{Understanding of optimal policy and good policy}
\label{app:opt-policy-good-policy}
In rollout policy and parallel rollout policy, we require to start with base policy $\pi.$ If the base policy is good then we can have good approximation for optimal value function. In this section, we understand about optimal policy and provide intuition for good policy.

%\subsection{Optimal policy} 
Let $\pi^{*}$ be the  optimal stationary policy. The expected discounted infinite horizon reward is higher than any other policy $\pi.$  
%That is, optimal policy $\pi^*$ consists of sequence of optimal actions under different states in different time. 
That is $\pi^* = \{\mu_1,\mu_2, \cdots \}.$ For all $\mathbf{x} \in \mathcal{S}$ the value function under optimal policy $\pi^* $ is 
\begin{eqnarray*}
	V_{\pi^*}(x) &=& E \left[ \sum_{t=1}^{\infty} \beta^{t}r(\mathbf{x}(t), \mu_t(\mathbf{x}(t)))~\vert~\mathbf{x}(1) = \mathbf{x}  \right] \\
	&=&
	E \left[ \sum_{t=1}^{\tau} \beta^{t}r(\mathbf{x}(t), \mu_t(\mathbf{x}(t)))~\vert~\mathbf{x}(1) = \mathbf{x}  \right] + 
	\\
	& & E \left[ \sum_{t=\tau}^{\infty} \beta^{t}r(\mathbf{x}(t), \mu_t(\mathbf{x}(t))) \right]
\end{eqnarray*}
We can bound second term in preceding eqn. when $r(\mathbf{x},a) \leq R_{\max}$ for all $\mathbf{x} \in \mathcal{S},$ and $a.$ Then we get  

\begin{eqnarray*}
	\bigg \vert E \left[ \sum_{t=\tau}^{\infty} \beta^{t}r(\mathbf{x}(t), \mu_t(\mathbf{x}(t))) \right] \bigg\vert
	\leq \frac{\beta^{\tau}R_{\max}}{1-\beta}.
\end{eqnarray*}
Thus, we can have 
\begin{eqnarray*}
	V_{\pi^*}(x) &\leq & 
	E \left[ \sum_{t=1}^{\tau} \beta^{t}r(\mathbf{x}(t), \mu_t(\mathbf{x}(t)))~\vert~\mathbf{x}(1) = \mathbf{x}  \right] +  \frac{\beta^{\tau}R_{\max}}{1-\beta}.
\end{eqnarray*}

For fixed $0<\beta<1,$ and $\epsilon > 0$ there exists $\tau < \infty$ such that  $\frac{\beta^{\tau}R_{\max}}{1-\beta} < \epsilon.$ This intuitively suggests that there is finite horizon $\tau$ from that reward is highest. Define 
\begin{eqnarray*}
	\widetilde{V}_{\pi^*,\tau} (\mathbf{x}) = E \left[ \sum_{t=1}^{\tau} \beta^{t}r(\mathbf{x}(t), \mu(\mathbf{x}(t)))~\vert~\mathbf{x}(1) = \mathbf{x}  \right]
\end{eqnarray*} 
This is finite horizon discounted value function, we can have $V_{\pi^*}(x) - \widetilde{V}_{\pi^*,\tau} (\mathbf{x}) < \epsilon.$ 

We generate sufficiently large number of trajectories $L$ using simulation model with policy $\pi^*,$ such that we can have following inequality.
{\small{
\begin{eqnarray*}
	\prob{ 
		\bigg \vert 
		\frac{1}{L} \sum_{l=1}^{L} \left( \sum_{t=1}^{\tau} \beta^{t-1} Q_{\pi^*,t}^{l}(\mathbf{x},\mu(\mathbf{x}),\mathbf{y}_{t,l}, \mu(\mathbf{y}_{t,l})) \right)
		-\widetilde{V}_{\pi^*,\tau}
		\bigg \vert 
		\geq  \epsilon 
	}   \\
	\leq  \delta 
\end{eqnarray*}
}}

Next using Hoeffding inequality, we can get the number of trajectories need to get desired level of accuracy. 

We next discuss good policy. Define
\begin{eqnarray*}
	\widetilde{\Pi} = \left\{ \pi \in \Pi: V_{\pi^*}(\mathbf{x}) \leq V_{\pi}(\mathbf{x}) + \epsilon  \right\}
\end{eqnarray*}
for all $\mathbf{x} \in \mathcal{S}.$
Thus $\widetilde{\Pi} $ is collection of policies which perform close to optimal policy, this is a set of good policy. Further, we can have 
\begin{eqnarray*}
	V_{\pi^*,\tau} (\mathbf{x})  - V_{\pi,\tau} (\mathbf{x}) \leq \epsilon
\end{eqnarray*}
for all $\pi \in \widetilde{\Pi}.$

%\subsection{title}
%\section{Error bounds for approximate value function and  policy iteration}

\end{document}